%% file: main.tex
\pdfoutput=1

\documentclass[english,a4paper,10pt]{article}

\usepackage[utf8]{inputenc}
\usepackage[T1]{fontenc}
\usepackage[english]{babel}

\usepackage{amsmath}
\usepackage{amssymb}
\usepackage{stmaryrd}
\usepackage{xargs}[2008/03/08]
\usepackage{xspace}
\usepackage[unicode=true,
            pdfusetitle,
            bookmarks=true,
            bookmarksnumbered=false,
            bookmarksopen=false,
            breaklinks=false,
            pdfborder={0 0 0},
            backref=false,
            colorlinks=true,
            citecolor=blue]
 {hyperref}

\usepackage{bussproofs}
\usepackage{centernot}
\usepackage{datetime}
\usepackage{filecontents}
\usepackage{fullpage}
\usepackage{lmodern}
\usepackage{microtype}
\usepackage{tikz}
\usepackage{pgfplots}
\usepackage{comment}
\usepackage{amsthm}
\usepackage[noabbrev,capitalize,nameinlink]{cleveref}

\usepackage{enumitem}
\setenumerate{itemsep=0.25em,topsep=0.25em,leftmargin=2em}
\setitemize{itemsep=0.25em,topsep=0.25em,leftmargin=2em}

\usetikzlibrary{automata}
\usetikzlibrary{backgrounds}
\usetikzlibrary{fit}
\usetikzlibrary{positioning}

\usepackage{listings}

\input{listing-style}

\usepackage{datetime}
\newdate{date}{08}{08}{2021}
\date{\displaydate{date}}

\begin{document}

\title{The SyGuS Language Standard Version 2.1}

\author{
  Saswat Padhi \and
  Elizabeth Polgreen \and
  Mukund Raghothaman \and
  Andrew Reynolds \and
  Abhishek Udupa
}

\maketitle

\input{macros}

\input{sygus-macros}

\section{Introduction}%
\label{sec:Introduction}

We present a language to specify instances of the syntax-guided synthesis
(SyGuS) problem. An instance of a SyGuS problem specifies:

\begin{enumerate}
\item
  The vocabularies, theories and the base types that are used to specify
  (a) semantic constraints on the function to be synthesized, and (b) the
  definition of the function to be synthesized itself. We refer to these as
  the \emph{input} and \emph{output} logics respectively.
\item
  A finite set of typed functions $f_1, \ldots, f_n$ that are to be synthesized.
\item
  The syntactic constraints on each function $f_i, i \in [1,n]$ to be
  synthesized. The syntactic constraints are specified using context-free
  grammars $G_i$ which describe the syntactic structure of candidate
  solution for each of these functions.
  The grammar may only involve function
  symbols and sorts from the specified output logic.
\item
  The semantic constraints and assumptions that describe the behavior of the functions to
  be synthesized. Constraints are given as quantifier-free formulas $\varphi$ 
  and assumptions are given as quantifier-free formulas $\alpha$
  in the input logic, and may refer to the functions-to-synthesize as well as 
  universally quantified variables
  $v_1, \ldots, v_m$. 
 
\end{enumerate}
The objective then is to find definitions $e_i$ (in the form of
expression bodies) for each function $f_i$ such that (a) the expression
body belongs to the grammar $G_i$ that is used to syntactically constrain
solutions for $f_i$, and 
(b) the constraint 
$\forall v_1, \ldots,
v_m \ldotp \alpha \implies \varphi$ is valid in the background theory
under the assumption that functions $f_1, \ldots, f_n$
are interpreted as functions returning $e_1, \ldots, e_n$.
Note that each $e_i$ is
an expression containing
a fixed set of free variables which
represent the argument list of the function $f_i$.


\paragraph{Overview of This Document}
This document defines the SyGuS format version 2.1,
which is intended to be used as the standard input and output language
for solvers targeting the syntax-guided synthesis problem.
The language borrows many concepts and language constructs
from the standard format for Satisfiability Modulo Theories (SMT)
solvers, the SMT-LIB standard (version 2.6)~\cite{BarFT-RR-17}.

\paragraph{Outline}
In the remainder of this section, we cover differences of
the SyGuS format described in this document with previous revisions~\cite{sygusFormat,sygusSyntax2015,sygusSyntax2016}
and cover the necessary preliminaries.
Then,
\cref{sec:syntax}
gives the concrete syntax for commands in the SyGuS input language.
\cref{sec:semantics}
documents the well-formedness and the semantics of input commands.
\cref{sec:output}
documents the expected output of synthesis solvers in response to these commands.
\cref{sec:sygus-logic}
describes formally the notion of a \emph{SyGuS logic} and how it restricts
the set of commands that are allowed in an input.
\cref{sec:logical-semantics}
formalizes what constitutes a correct response to a SyGuS input.
Finally, \cref{sec:examples}
provides examples of possible inputs in the SyGuS language
and solvers responses to them.

\subsection{Differences from Previous Versions}

In this section, we cover the differences
in this format with respect to the one described in the previous
version of the SyGuS format~\cite{sygusFormat},
and its extensions~\cite{sygusSyntax2015,sygusSyntax2016}.
\subsubsection{Changes from SyGuS 1.0 to SyGuS 2.0}
\begin{enumerate}

\item The syntax for providing grammars inside the $\synthfunkwd$ command
now requires that non-terminal symbols are declared upfront
in a \emph{predeclaration}, see \cref{ssec:dec-synth-fun}
for details.

\item 
The keyword \code|Start|, which 
denoted the starting (non-terminal) symbol of
grammars in the previous standard, has been removed.
Instead, the first symbol listed in the grammar is assumed
to be the starting symbol.

\item 
Terms that occur as the right hand side of production rules in
SyGuS grammars are now required to be binder-free.
In particular, this means that let-terms are now disallowed within grammars.
Accordingly,
the keywords $\inputvarkwd$ and $\localvarkwd$,
which were used to specify input and local variables in grammars
respectively, have been removed since the former is equivalent to $\varkwd$
and the latter has no affect on the grammar.

\item The datatype keyword $\enumkwd$
and related syntactic features have been removed.
The standard SMT-LIB 2.6 commands
for declaring datatypes are now adopted,
see \cref{ssec:declaring-sorts} for details.

\item The $\setoptskwd$ command has been renamed $\setoptkwd$ to 
correlate to the existing SMT-LIB version 2.6 command.

\item 
The syntax for terms and sorts
now coincides with the corresponding syntax for terms and sorts 
from SMT-LIB versions 2.0 and later.
There are three notable changes
with respect to the previous SyGuS format that come as a result of this change.
First,
negative integer and real constants must be written with
unary negation, that is, the integer constant negative one
must be written \code|(- 1)|, whereas previously it could
be written \code|-1|.
Second,
the syntax for bit-vectors sorts \code|(BitVec n)|
is now written  \code|(_ BitVec n)|.
Third,
all let-bindings do \emph{not} annotate the type of the variable being bound.
Previously, a let-term was written \code|(let ((x T t)) ...)|
where \code|T| indicates the type of \code|t|.
Now, it must be written in the SMT-LIB compliant
way \code|(let ((x t)) ...)|.

\item
The signature, syntax and semantics for the theory of strings
is now the one given in an initial proposal~\cite{smtlibStrings} to SMT-LIB.
Thus, certain new symbols are now present in the signature,
and some existing ones have changed names.
The semantics of all operators,
which was not specified previously,
is now consistent with that provided in the proposal~\cite{smtlibStrings}.


\item The command $\primedvardeclkwd$, which was a syntactic
sugar for two $\vardeclkwd$ commands in the previous language standard, has been removed
for the sake of simplicity.
This command did not provide any benefit to
invariant synthesis problems, since constraints declared via $\constraintinvkwd$
do not accept global variables.



\item 
The command $\fundeclkwd$,
which declared a universal variable of function type
in the previous language standard,
has been removed.

\item
A formal notion of \emph{input logics}, \emph{output logics}, and 
\emph{features} have been defined as part of the SyGuS background logic,
and the command $\setfeaturekwd$ 
has been added to the language,
which is used for further refining
the constraints, grammars and commands that may appear in an input.

\item The expected output for fail responses from a solver has changed.
In particular,
the response \code|(fail)| from the previous language standard
should now be provided as \code|fail|, that is, without parentheses.
Additionally, a solver may answer \code|infeasible|,
indicating that it has determined there are no solutions to the given conjecture.

\end{enumerate}
\subsubsection{Changes from SyGuS 2.0 to SyGuS 2.1}
\begin{enumerate}
\item We introduce the command $\assumekwd$, which introduces a list of assumptions $\alpha$, and the 
synthesis conjecture is now 
$\exists f_1, \ldots, f_n \ldotp \forall v_1, \ldots, v_m \ldotp \alpha \implies \varphi$. 

\item Oracle constraints are introduced. Oracle constraints are constraints that are generated when an external oracle is called with a concrete set of input values. The constraints are conjoined to the list of constraints $\varphi$ in the synthesis conjecture.

\item Oracle assumptions are introduced. Oracle assumptions are assumptions that are generated when an external oracle is called with a concrete set of input values. The assumptions are conjoined to the list of assumptions $\alpha$ in the synthesis conjecture.

\item Oracle functions are introduced. These are functional symbols whose interpretation is associated to an external oracle. The functions are treated as universally quantified uninterpreted functions, and calls to that external oracle place with concrete input values generate assumptions over the behavior of the uninterpreted function. These assumptions are conjoined to the list of assumptions $\alpha$ in the synthesis conjecture.

\item A set of new commands are introduced to enable easy specification of specific types of oracles, for instance input-output oracles. These oracles are based on the formalization by Jha and Seshia~\cite{oracles}.

\item A new theory of tables is referenced as a possible underlying theory in \cref{ssec:smt-logic}.

\item A new SyGuS logic, called \code|CHC_$X$|,
has been introduced for specifying synthesis problems using constrained Horn clause systems
over an SMT-LIB base logic $X$.

\item A new command, $\constraintchckwd$, has been introduced for specifying constraints in the form of constrained Horn clauses.

\item The $\synthinvkwd$ command, used within invariant synthesis (\code|Inv_$X$|) logic earlier, has been removed.
Instead, all functions (including predicates) to be synthesized are now specified using $\synthfunkwd$.

\item The expected successful response for SyGuS solvers to a $\checksynthkwd$ command is now enclosed in parentheses.

\item A new command $\weightdeclkwd$, has been introduced for introducing user-defined weight attributes.

\item A new feature $\fweightskwd$ has been added, which allows weight annotations on terms, 
as well as the use of weight symbols within terms, as described in \cref{sec:weight-semantics}.

\item A new command $\optsynthkwd$ has been added for specifying optimization queries.

\item
The signature, syntax and semantics for the theory of strings
is now the one given in the official version 2.6 release of the theory of Unicode strings~\cite{BarFT-RR-17}.
\end{enumerate}

\subsection{Preliminaries}
In this document, we assume basic standard notions of
multi-sorted first-order logic.
We assume the reader is familiar
with sorts, well-sorted terms, (quantified) formulas
and free variables\footnote{
The definition of each of these coincides with
the definition given in the SMT-LIB 2.6 standard~\cite{BarFT-RR-17}.
}
If $e$ is a term or formula,
then we write $e[x]$ to denote that $x$ occurs free in $e$,
and $e[t]$ to denote the result of replacing all occurrences of $x$ by $t$.
We write $\lambda \vec{x} \ldotp t[\vec{x}]$ to denote a \emph{lambda term},
that is, an anonymous function whose argument list is $\vec{x}$
that returns the value of $t[\vec{s}]$ for all inputs $\vec{x} = \vec{s}$.
Given an application of a lambda term to a concrete argument list $\vec{s}$,
i.e. the term $(\lambda \vec{x} \ldotp t[\vec{x}( \vec{s} )$,
then its \emph{beta-reduction} is the term $t[\vec{s}]$.
We use $\teq$ to denote the binary (infix) equality predicate.

\section{Syntax}
\label{sec:syntax}

In this section, we describe the concrete syntax
for the SyGuS 2.0 input language.
Many constructs in this syntax coincide
with those in SMT-LIB 2.6 standard~\cite{BarFT-RR-17}.
In the following description,
italic text within angle-brackets represents EBNF non-terminals,
and text in typewriter font represents terminal symbols.

A SyGuS input $\sygus$ is thus a sequence of zero or more commands.
\[
\begin{array}{rcl}
\sygus & ::= & \kstar{\cmd}\\[2ex]
\end{array}
\]
We first introduce the necessary preliminary definitions,
and then provide the syntax for commands $\cmd$ at the end of this section.

\subsection{Comments}

Comments in SyGuS specifications are indicated by a semicolon ${\tt ;}$.
After encountering a semicolon, the rest of the line is ignored.

\subsection{Literals}
\label{ssec:literals}

A \emph{literal} $\literal$ is a special sequence of characters,
typically used to denote values or constant terms.
The SyGuS format includes syntax for several kinds of literals,
which are listed below.
This treatment of most of these literals coincides with
those in SMT-LIB version 2.6.
For full details, see Section 3.1 of the SMT-LIB 2.6 standard~\cite{BarFT-RR-17}.

\begin{alignat*}{1}
 & \begin{array}{rcl}
\literal & ::= & \begin{array}{ccccccccccc}
\intconst & | & \realconst & | & \boolconst & | \\
\hexconst & | & \binaryconst & | & \stringconst\end{array}\\
\end{array}
\end{alignat*}

\paragraph{Numerals ($\intconst$)}
Numerals are
either the digit $0$,
or a non-empty sequence of digits $\left[{\tt 0}-{\tt 9}\right]$
that does not begin with $0$.

\paragraph{Decimals ($\realconst$)}
The syntax for decimal numbers is $\intconst{\tt .\kstar{0}}\intconst$.

\paragraph{Booleans ($\boolconst$)}
Symbols $\truekwd$ and $\falsekwd$ denote the Booleans true and false.

\paragraph{Hexidecimals ($\hexconst$)}
Hexadecimals are written with ${\tt \#x}$
followed by a non-empty sequence of (case-insensitive) 
digits and letters taken from the ranges $\left[{\tt A}-{\tt F}\right]$
and $\left[{\tt 0}-{\tt 9}\right]$.

\paragraph{Binaries ($\binaryconst$)}
Binaries are written with ${\tt \#b}$
followed by a non-empty sequence of bits $\left[{\tt 0}-{\tt 1}\right]$.

\paragraph{Strings ($\stringconst$)}
A string literal $\stringconst$
is any sequence of printable characters
delimited by double quotes $\texttt{"}$ $\texttt{"}$.
The characters within these delimiters
are interpreted as denoting characters of the string in a one-to-one correspondence,
with one exception:
two consecutive double quotes within a string
denote a single double quotes character.
In other words, \code|"a""b"| denotes the string
whose characters in order are \code|a|, \code|"| and \code|b|.
Strings such as \code|"\n"| whose characters are commonly
interpreted as escape sequences are not handled specially,
meaning this string is interpreted 
as the one consisting of two characters, \code|\| followed by \code|n|.

\noindent
Literals are commonly
used for denoting $0$-ary symbols of a theory.
For example, 
the theory of integer arithmetic 
uses numerals to denote non-negative integer values.
The theory of bit-vectors uses both
hexadecimal and binary constants in the above syntax
to denote bit-vector values.

\subsection{Symbols}

Symbols are denoted with the non-terminal $\symbol$. 
A symbol
is any non-empty sequence of upper- and lower-case alphabets, digits,
and certain special characters (listed below), with the restriction that it may not
begin with a digit and is not a reserved word (see \cref{apx:reserved} 
for a full list of reserved words).
A special character is any of the following:
\begin{center}
  \texttt{\_ + - * \& | ! \string~ < > = / \% ? . \$ \string^}
\end{center}
Note this definition coincides with simple
symbols in Section 3.1 of SMT-LIB version 2.6,
apart from differences in their reserved words.

\paragraph{Keywords ($\keyword$)}
Following SMT-LIB version 2.6,
a keyword $\keyword$ is a symbol whose first character is \code|:|.

\subsection{Identifiers}

An identifier $\identifier$
is a syntactic extension of symbols 
that includes symbols that are indexed by integer constants or other symbols.
\[
\begin{array}{rcl}
\identifier & ::= & \begin{array}{ccc}
\symbol & | & \paren{{\tt \_}\mbox{ }\symbol\mbox{ }\kplus{\identifierindex}}
\end{array}\\
\identifierindex & ::= & \begin{array}{ccc}
\intconst & | & \symbol
\end{array}\\
\end{array}
\]
Note this definition coincides with
identifiers in Section 3.1 of SMT-LIB version 2.6.

\subsection{Attributes}

An attribute $\attribute$
is a keyword and an (optional) associated value.
\[
\begin{array}{rcl}
\attribute & ::= & \begin{array}{ccc}
\keyword & | & \keyword \attributevalue
\end{array}\\
\end{array}
\]
The permitted values of attribute values $\attributevalue$ depend on the attribute
they are associated with.
Possible values of attributes include symbols, as well as (lists of) sorts and terms.
The above definition of attribute coincides with
attributes in Section 3.4 of SMT-LIB version 2.6.
All attributes standardized in this document are listed in \cref{sec:attr}.

\subsubsection{Weight Attributes}
\label{sec:weight-attributes}
Beyond the SMT-LIB standard, we standardize one class of attributes, namely that of \emph{weights}.
Some keywords we classify as \emph{weight keywords}.
In particular, we assume that \kweight\ is the (builtin) weight keyword.
Additional weight keywords can be user-defined via the command $\weightdeclkwd$.

A weight attribute consists of a weight keyword and an attribute value that is a integer numeral.
For example, $\kweight\mbox{ }5$ denotes the attribute associating the builtin weight keyword to $5$.
Terms can be annotated with weight attributes (see \cref{ssec:term-annotations}),
which will have a special semantics which we will describe in detail in \cref{sec:weight-semantics}.

\subsection{Sorts}

We work in a multi-sorted logic where terms 
are associated with sorts $\sortexpr$.
Sorts are constructed via the following syntax.
\begin{alignat*}{1}
 & \begin{array}{rcl}
\sortexpr & ::= & \identifier\mbox{ }|\mbox{ }\paren{\identifier\mbox{ }\kplus{\sortexpr}}\\
\end{array}
\end{alignat*}
The \emph{arity} of the sort is the number of (sort) arguments it takes.
A \emph{parametric} sort is one whose arity is greater than zero.
Theories associate identifiers with sorts and sort constructors
that have an intended semantics.
Sorts may be defined by theories (see \cref{ssec:smt-logic})
or may be user-defined
(see \cref{ssec:declaring-sorts}).

\subsection{Terms}

We use terms $\term$ to specify grammars and constraints,
which are constructed by the following syntax.
\begin{alignat*}{1}
 & \begin{array}{rcl}
\term & ::= & \identifier\\
 & | & \literal\\
 & | & \paren{\identifier\mbox{ }\kplus{\term}}\\
 & | & \paren{!\mbox{ }\term\mbox{ }\kplus{\attribute}}\\
 & | & \paren{\existskwd\mbox{ }\paren{\kplus{\sortedvar}}\mbox{ }\term}\\
 & | & \paren{\forallkwd\mbox{ }\paren{\kplus{\sortedvar}}\mbox{ }\term}\\
 & | & \paren{\letkwd\mbox{ }\paren{\kplus{\varbinding}}\mbox{ }\term}\\[2ex]
 \bfterm & ::= & \identifier\\
 & | & \literal\\
 & | & \paren{\identifier\mbox{ }\kplus{\bfterm}}\\
 & | & \paren{!\mbox{ }\bfterm\mbox{ }\kplus{\attribute}}\\[2ex]
 \sortedvar & ::= & \paren{\symbol\mbox{ }\sortexpr}\\
 \varbinding & ::= & \paren{\symbol\mbox{ }\term}\\
\end{array}
\end{alignat*}
Above,
we distinguish a subclass of \emph{binder-free} terms $\bfterm$ in the syntax above,
which do not contain bound (local) variables.
Like sorts, the identifiers that comprise terms
can either be defined by the user or by background theories.

\subsubsection{Term Annotations}
\label{ssec:term-annotations}
In SMT-LIB,
terms $t$ may be annotated with \emph{attributes}.
The purpose of an attribute is to mark a term with a set of special properties, 
which may influence the expected result of later commands.
Attributes are specified using the
syntax $\paren{!\mbox{ }t\mbox{ }A_1\mbox{ }\ldots\mbox{ }A_n}$
where $t$ is a term and $A_1, \ldots, A_n$ are attributes.
An attribute can be any
The term above is semantically equivalent to $t$ itself.
Several attributes are standardized by the SMT-LIB standard, while
others may be user-defined.

\subsection{Features}
\label{ssec:syntax-features}

A feature $\feature$ is a keyword denoting a restriction or extension
on the kinds of SyGuS commands that are allowed in an input.
It is an enumeration in the following syntax.
\begin{alignat*}{1}
 & \begin{array}{rcl}
 \feature & ::= & :\fgrammarskwd\mbox{ }|\mbox{ }:\ffwddeclskwd\mbox{ }|\mbox{ }:\frecursionkwd\mbox{ }|\mbox{ } :\foracleskwd\mbox{ }|\mbox{ } :\fweightskwd\mbox{ }
\end{array}
\end{alignat*}
More details on features are given in \cref{ssec:feature-sets}.

\subsection{Commands}

A command $\cmd$ is given by the following syntax.

\[
\begin{array}{rcl}
\cmd 
 & ::= & \paren{\assumekwd\mbox{ }\term} \\
 & | & \paren{\checksynthkwd} \\
 & | & \paren{\constraintchckwd\mbox{ }\paren{\kstar{\sortedvar}}\mbox{ }\term\mbox{ }\term} \\
 & | & \paren{\constraintkwd\mbox{ }\term} \\
 & | & \paren{\vardeclkwd\mbox{ }\symbol\mbox{ }\sortexpr}\\ 
 & | & \paren{\weightdeclkwd\mbox{ }\symbol\mbox{ }\kstar{\attribute}} \\
 & | & \paren{\constraintinvkwd\mbox{ }\symbol\mbox{ }\symbol\mbox{ }\symbol\mbox{ }\symbol} \\
 & | & \paren{\optsynthkwd\mbox{ }\paren{\kstar{\term}}\mbox{ }\kstar{\attribute}} \\
 & | & \paren{\setfeaturekwd\mbox{ }\feature\mbox{ }\boolconst} \\
 & | & \paren{\synthfunkwd\mbox{ }\symbol\mbox{ }\paren{\kstar{\sortedvar}}\mbox{ }\sortexpr\mbox{ }\koption{\grammardef}}\\
 & | & \oraclecmd \\
 & | & \smtcmd \\[2ex]

 \oraclecmd 
 & ::= & \paren{\assumeoraclekwd\mbox{ }\paren{\kstar{\sortedvar}}\mbox{ }\paren{\kstar{\sortedvar}}\mbox{ }
  \term \mbox{ } \symbol\mbox{ }} \\
  & | & \paren{\constraintoraclekwd\mbox{ }\paren{\kstar{\sortedvar}}\mbox{ }\paren{\kstar{\sortedvar}}\mbox{ }
  \term\mbox{ } \symbol\mbox{ }} \\
 & | & \paren{\oracledeckwd\mbox{ }\symbol\mbox{ }\paren{\kstar{\sortexpr}}\mbox{ }\sortexpr\mbox{ }\symbol\mbox{ }} \\
 & | &\paren{\iooracledeckwd\mbox{ }\symbol\mbox{ } \symbol\mbox{ }} \\
 & | &\paren{\cexoracledeckwd\mbox{ }\symbol\mbox{ } \symbol\mbox{ }} \\
 & | &\paren{\memoracledeckwd\mbox{ }\symbol\mbox{ } \symbol\mbox{ }} \\
 & | &\paren{\poswitnessoracledeckwd\mbox{ }\symbol\mbox{ } \symbol\mbox{ }} \\
 & | &\paren{\negwitnessoracledeckwd\mbox{ }\symbol\mbox{ } \symbol\mbox{ }} \\
 & | &\paren{\corroracledeckwd\mbox{ }\symbol\mbox{ } \symbol\mbox{ }} \\
 & | &\paren{\corrcexoracledeckwd\mbox{ }\symbol\mbox{ } \symbol\mbox{ }} \\[2ex]

 \smtcmd 
 & ::= &\paren{\dtdeclkwd\mbox{ }\symbol\mbox{ }\dtdecl} \\
 & | & \paren{\dtsdeclkwd\mbox{ }\paren{\kstarn{\sortdecl}{n+1} }\mbox{ }\paren{\kstarn{\dtdecl}{n+1}}} \\
 & | & \paren{\sortdeclkwd\mbox{ }\symbol\mbox{ }\intconst} \\
 & | & \paren{\fundefkwd\mbox{ }\symbol\mbox{ }\paren{\kstar{\sortedvar}}\mbox{ }\sortexpr\mbox{ }\term} \\  
 & | & \paren{\sortdefkwd\mbox{ }\symbol\mbox{ }\sortexpr} \\
 & | & \paren{\setinfokwd\mbox{ }\keyword{ }\literal} \\
 & | & \paren{\setlogickwd\mbox{ }\symbol} \\
 & | & \paren{\setoptkwd\mbox{ }\keyword\mbox{ }\literal}
\end{array}
\]

\[
\begin{array}{rcl}  
 \sortdecl& ::= & \paren{\symbol\mbox{ }\intconst}\\
 \dtdecl & ::= & \paren{\kplus{\dtconsdecl}} \\
 \dtconsdecl & ::= & \paren{\symbol\mbox{ }\kstar{\sortedvar}} \\[2ex]
 \grammardef & ::= & \paren{\kstarn{\sortedvar}{n+1}}\mbox{ }\paren{\kstarn{\ntdef}{n+1}} \\
 \ntdef & ::= & \paren{\symbol\mbox{ }\sortexpr\mbox{ }\paren{ \kplus{\gterm} } } \\[2ex]
 \gterm 
 & ::= &  \paren{\constantkwd\mbox{ }\sortexpr}\mbox{ }|\mbox{ }\paren{\varkwd\mbox{ }\sortexpr}\mbox{ }|\mbox{ }\bfterm \\[2ex]
\end{array}
\]
For convenience,
we distinguish between three kinds of commands above.
The commands listed under $\cmd$ and $\oraclecmd$ are specific to the SyGuS format,
with the latter pertaining to oracles.
The remaining commands listed under $\smtcmd$
are borrowed from SMT-LIB 2.6.
The semantics of these commands are detailed in \cref{sec:semantics}.


\section{Semantics of Commands}%
\label{sec:semantics}

A SyGuS input file is a sequence of commands,
which at a high level
are used for defining a (single) synthesis conjecture,
and invoking a solver for this conjecture.
This conjecture is a closed formula of the form:
\begin{alignat*}{1}
 & \exists f_1,\ldots,f_n \ldotp \forall v_1,\ldots,v_m \ldotp (\alpha_1 \wedge \ldots \wedge \alpha_r) \implies (\varphi_1 \wedge \ldots \wedge \varphi_q )
\end{alignat*}
We wil use $\Psi$ to refer to the formula $\forall v_1,\ldots,v_m \ldotp (\alpha_1 \wedge \ldots \wedge \alpha_r)\implies (\varphi_1 \wedge \ldots \wedge \varphi_q )$, and thus the synthesis conjecture can be written $\exists f_1,\ldots,f_n \ldotp \Psi$. 
In this section, we define how this conjecture is
is established via SyGuS commands.
Given a sequence of commands, the current state consists of the following
information:
\begin{itemize}
\item A list
$f_1, \ldots, f_n$, which we refer to
as the current list of \emph{functions to synthesize},
\item A list 
$v_1, \ldots, v_m$ of variables, 
which we refer to as the current list of \emph{universal variables},
\item A list of formulas
$\varphi = \varphi_1, \ldots, \varphi_q$,
which we refer to as the current list of \emph{constraints},
\item A list of formulas
$\alpha = \alpha_1, \ldots, \alpha_r$,
which we refer to as the current list of \emph{assumptions},
\item A \emph{signature}
denoting the set of defined symbols in the current scope.
A signature is 
a mapping from symbols to expressions (either sorts or terms).
Each of these symbols may have a predefined semantics
either given by the theory,
or defined by the user (e.g. symbols that are defined as macros
fit the latter category).
\item A \emph{SyGuS logic} denoting the
terms and sorts that may appear in constraints and grammars.
\end{itemize}
In the initial state of a SyGuS input,
the lists of functions-to-synthesize, 
universal variables, constraints, assumptions, and the signature are empty,
and the SyGuS logic is the default one 
(see \cref{sec:sygus-logic} for details).  

In the following, we first describe restrictions
on the order in which commands can be specified in SyGuS inputs.
We then describe how each command $\cmd$ updates
the state of the sets above and the current signature.

\subsection{Command Ordering}
A SyGuS input is not well-formed
if it specifies a list of commands that do not meet
the restrictions given in this section regarding their order. The
order is specified by the following regular pattern:

\vspace*{2mm}
\noindent
\begin{equation*}
(\mathit{\{ set\ logic\ command \}})?\mbox{ }\mathit{(\{setter\ commands\})^*\mbox{ }(\{other\ commands\})^*}
\end{equation*}
where the set $\mathit{\{set\ logic\ command\}}$ consists
of the set of all $\setlogickwd$ commands,
set $\mathit{\{setter\ commands\}}$ includes
the \texttt{set-feature} and \texttt{set-option}
commands, the set $\mathit{\{other\ commands\}}$ include all the SyGuS
commands except the commands in the aforementioned two sets.

In other words, a SyGuS input is well formed if it begins with 
at most one \texttt{set-logic} command, followed by 
a block of zero or more \texttt{set-feature}
and \texttt{set-option} commands in any order, 
followed by zero or more instances of the other SyGuS commands.

\subsection{Setting the Logic}%
\label{ssec:set-logic}

The logic of a SyGuS specification consists of three parts ---
an \emph{input logic}, an \emph{output logic} and a \emph{feature set}.
Roughly, 
the input logic determines what terms can appear in constraints,
and the output logic determines what terms can appear in grammars and solutions.
The feature set places additional restrictions or extensions
on the constraints, grammars as well as the commands that are allowed in an input.
These are described in detail in \cref{sec:sygus-logic}.

\begin{itemize}
\item $\paren{\setlogickwd\mbox{ }L}$

This command sets the SyGuS background logic to the one that $L$ refers to.
The logic string $L$ can be a standard one defined in the SMT-LIB 2.6 standard~\cite{BarFT-RR-17}
or may be solver-specific.
If $L$ is an SMT-LIB standard logic,
then it must contain quantifiers or this command is not well-formed,
that is, logics with the prefix \code|QF_| are not allowed.\footnote{
By convention quantifiers are always included in the logic. This is because
the overall synthesis conjecture specified by the
state may involve universal quantification.
}
If this command is well-formed and $L$ is an SMT-LIB standard logic,
then this command sets the SyGuS logic to the one whose
input and output logics are \code|QF_$L$| and whose feature
set is the default one defined in this document (\cref{ssec:feature-sets}).
In other words, when this command has $L$ as an argument
and $L$ is an SMT-LIB standard logic,
this indicates that that terms in the logic of $L$ are
allowed in constraints, grammars and solutions,
but they are are restricted to be quantifier-free.
As a consequence,  the overall synthesis conjectures 
allowed by default when $L$ is a standard SMT-LIB logic
have at most two levels of quantifier alternation.

\item $\paren{\setfeaturekwd\mbox{ }:F\mbox{ }b}$

This command enables the feature specified by $F$
in the feature set component of the SyGuS background logic
if $b$ is $\truekwd$, or disables it if $b$ is $\falsekwd$.
All features standardized in SyGuS 2.0
are given in \cref{ssec:feature-sets}.

\end{itemize}

\subsection{Declaring Universal Variables}

\begin{itemize}
\item $\paren{\vardeclkwd\mbox{ }S\mbox{ }\sigma}$

This command appends $S$ to the current list of universal variables
and adds the symbol $S$ of sort $\sigma$ to the current signature.
This command should be rejected if $S$ already
exists in the current signature.





\end{itemize}

\subsection{Declaring Functions-to-Synthesize}
\label{ssec:dec-synth-fun}

\begin{itemize}
\item $\paren{\synthfunkwd\mbox{ }S\mbox{ }
\paren{\paren{x_1\mbox{ }\sigma_1} \ldots \paren{x_n\mbox{ }\sigma_n} }\mbox{ }\sigma\mbox{ }
\koption{G}}$

This command adds $S$ to the current list of functions to synthesize,
and adds the symbol $S$ of 
sort $\sigma_1 \times \ldots \times \sigma_n \rightarrow \sigma$
to the current signature.
This command should be rejected if $S$ is already
a symbol in the current signature.
We describe restrictions and well-formedness requirements for this command
in the following.

If provided, the syntax for the grammar $G$
consists of two parts: 
a \emph{predeclaration}
$\paren{\paren{y_1\mbox{ }\tau_1} \ldots \paren{y_n\mbox{ }\tau_n}}$,
followed by a \emph{grouped rule listing}
$\paren{\paren{y_1\mbox{ }\tau_1\mbox{ }\paren{g_{11} \ldots g_{1m_1} } } \ldots \paren{y_n\mbox{ }\tau_n\mbox{ }\paren{g_{n1} \ldots g_{nm_n} } }}$
where $y_1, \ldots, y_n$ are the \emph{non-terminal} symbols of the grammar.
Note that the same variable symbols
$y_1, \ldots, y_n$ and their sorts 
$\tau_1, \ldots, \tau_n$ appear both in the predeclaration and as heads
of each of the rules.
If this is not the case, then this command is not well-formed.
For all $i,j$, recall that grammar term $g_{ij}$ is either a term, or a class of terms
denoted by $\paren{\constantkwd\mbox{ }\sigma_c}$ 
and $\paren{\varkwd\mbox{ }\sigma_v}$ denoting respectively
the set of constants whose sort is $\sigma_c$,
and the set of all variables from $x_1, \ldots, x_n$ whose sort is $\sigma_v$.
If $g_{ij}$ is an ordinary term, then its free variables may contain $y_1, \ldots, y_n$,
as well as $S$ itself.
If the grammar contains $S$ itself, then it is possible that the definition
given for $S$ in a solution is recursive, however,
this feature is disallowed in the default logic (see \cref{sec:sygus-logic}).

This command is not well-formed if
$\tau_1$ (the type of the start symbol) is not $\sigma$.
It is also not well-formed if
$G$ generates a term $t$ from $y_i$
that does not have type $\tau_i$ for some $i$.

If provided, the grammar $G$ must also be one that is allowed
by the output logic of the current SyGuS logic.
For more details on the restrictions imposed on grammars by the logic,
see \cref{sec:sygus-logic}.
If $G$ does not meet the restrictions
of the background logic, it should be rejected.

If no grammar is provided,
then any term of the appropriate sort
in the output logic may be given in the body of a solution for $S$.

More details on grammars and the terms
they generate, as well as what denotes a term that meets the syntactic
restrictions of a function-to-synthesize
is discussed in detail in \cref{ssec:sat-syntactic}.

\end{itemize}

\subsection{Declaring Sorts}
\label{ssec:declaring-sorts}

In certain logics, 
it is possible for the user to declare user-defined sorts.
For example, 
$\dtsdeclkwd$ commands may be given
when the theory of datatypes is enabled in the background
logic,
$\sortdeclkwd$ commands may be given
when uninterpreted sorts are enabled in the background logic.

\begin{itemize}
\item $\paren{\dtdeclkwd\mbox{ }S\mbox{ }D}$

This command is syntax sugar for
$\paren{\dtsdeclkwd\mbox{ }\paren{\paren{S\mbox{ }0} }\mbox{ }\paren{D}}$.

\item $\paren{\dtsdeclkwd\mbox{ }\paren{\paren{S_1\mbox{ }a_1}\ldots\paren{S_n\mbox{ }a_n} }\mbox{ }\paren{D_1 \ldots D_n}}$

This command adds symbols corresponding to the datatype
definitions $D_1, \ldots, D_n$ for $S_1, \ldots, S_n$ 
to the current signature.
For each $i = 1, \ldots, n$, integer constant $a_i$ denotes the arity of 
datatype $S_i$.
The syntax of 
$D_i$ is a \emph{constructor listing} of the form
\[
\paren{
\paren{c_1\mbox{ }\paren{s_{11}\mbox{ }\sigma_{11}}\mbox{ }\ldots\mbox{ }\paren{s_{1m_1}\mbox{ }\sigma_{1m_1}} }
\mbox{ }\ldots\mbox{ }
\paren{c_k\mbox{ }\paren{s_{k1}\mbox{ }\sigma_{k1}}\mbox{ }\ldots\mbox{ }\paren{s_{km_k}\mbox{ }\sigma_{km_k}} }
}
\]
For each $i$, the following symbols are added to the signature:
\begin{enumerate}
\item 
Symbol $S_i$ is added to the current signature,
defined it as a datatype sort whose definition is given by $D_i$,
\item 
Symbols $c_1, \ldots, c_k$ are added to the signature,
where
for each $j = 1, \ldots, k$, symbol $c_j$
is defined as a \emph{constructor}
of sort $\sigma_{j1} \times \ldots \times \sigma_{jm_j} \rightarrow D_i$,
\item 
For each $j = 1, \ldots, k$, $\ell = 1, \ldots m_j$, symbol $s_{j\ell}$
is added to the signature,
defined as a \emph{selector}
of sort $D_i \rightarrow \sigma_{j\ell}$.
\end{enumerate}

This command should be rejected if any of the above symbols this command
adds to the signature are already
a symbol in the current signature.
We provide examples of datatype definitions in \cref{sec:examples}.
For full details on well-formed datatype declarations,
refer to Section 4.2.3 of the SMT-LIB 2.6 standard~\cite{BarFT-RR-17}.

\item $\paren{\sortdeclkwd\mbox{ }S\mbox{ }n}$

This command adds the symbol $S$ to the current signature
and associates it with an uninterpreted sort of arity $n$.
This command should be rejected if $S$ is already
a symbol in the current signature.

\end{itemize}

\subsection{Declaring Weight Keywords}
\label{ssec:declaring-weights}

\begin{itemize}
\item $\paren{\weightdeclkwd\mbox{ }S\mbox{ }A_1 \ldots A_n}$

This command declares the symbol $:S$ as a weight keyword.
The attributes $A_1 \ldots A_n$ can be used to indicate properties of the weight,
such as its default value.
For details, see \cref{sec:attr-weights}.

Terms can be subsequently annotated
with this keyword, e.g. $\paren{!\mbox{ }t\mbox{ }:S\mbox{ }1}$,
which are then given a special semantics as described in \cref{sec:weight-semantics}.

\end{itemize}
This command is only allowed when the $\fweightskwd$ feature is enabled.

\subsection{Defining Macros}

\begin{itemize}
\item $\paren{\fundefkwd\mbox{ }S\mbox{ }\paren{\paren{x_1\mbox{ }\sigma_1} \ldots \paren{x_n\mbox{ }\sigma_n}}\mbox{ }\sigma\mbox{ }t}$

This command adds to the current signature
the symbol $S$ of sort $\sigma$
if $n=0$ or $\sigma_1 \times \ldots \sigma_n \rightarrow \sigma$ if $n>0$.
The variables $x_1, \ldots, x_n$ may occur freely in $t$.
It defines $S$ as a term whose semantics are given by the function
$\lambda x_1, \ldots, x_n \ldotp t$.
Notice that $t$ may not contain any free occurrences of $S$,
that is, the definition above is not recursive.
This command is not well-formed if $t$ is not a well-sorted
term of sort $\sigma$.
This command should be rejected if $S$ is already
a symbol in the current signature.

\item $\paren{\sortdefkwd\mbox{ }S\mbox{ }\paren{u_1 \ldots u_n}\mbox{ }\sigma}$

This command adds the symbol $S$ to the current signature.
It defines $S$ as the sort $\sigma$.
The sort variables $u_1, \ldots, u_n$
may occur free in $\sigma$,
while $S$ may not occur free in $\sigma$.
This command is not well-formed if $\sigma$
is not a well-formed sort.
This command should be rejected if $S$ is already
a symbol in the current signature.

\end{itemize}

\subsection{Asserting Synthesis Constraints and Assumptions}%
\label{ssec:asserting}

\begin{itemize}
\item $\paren{\constraintkwd\mbox{ }t}$

This command adds $t$ to the set of constraints.
This command is well formed if $t$ is a well-sorted formula,
that is, a well-sorted term of sort $\sbool$.
Furthermore,
the term $t$ should be allowed
based on the restrictions of the current logic,
see \cref{sec:sygus-logic} for more details.

\item $\paren{\assumekwd\mbox{ }t}$

This command adds $t$ to the set of assumptions.
Like constraints,
this command is well formed if $t$ is a well-sorted term of sort $\sbool$.
and is allowed based on the restrictions of the current logic.

\item $\paren{\constraintinvkwd\mbox{ }S\mbox{ }S_{pre}\mbox{ }S_{trans}\mbox{ }S_{post}}$

This command adds a set of constraints to the current 
state that correspond to an invariant synthesis problem for function-to-synthesize $S$,
where $S_{pre}$ denotes a pre-condition,
$S_{post}$ denotes a post-condition
and $S_{trans}$ denotes a transition relation.

A constraint of this form is well-formed if:
\begin{enumerate}
\item
$S$ is the function-to-synthesize
of sort $\sigma_1 \times \ldots \times \sigma_n \rightarrow \sbool$,
\item
$S_{pre}$ is a defined symbol 
whose definition is of the form $\lambda x_1, \ldots, x_n \ldotp \varphi_{pre}$,
\item
$S_{trans}$ is a defined symbol 
whose definition is of the form $\lambda x_1, \ldots, x_n, y_1, \ldots, y_n \ldotp \varphi_{trans}$, and
\item
$S_{post}$ is a defined symbol 
whose definition is of the form $\lambda x_1, \ldots, x_n \ldotp \varphi_{post}$.
\end{enumerate}
where $(x_1, \ldots, x_n)$ and $(y_1, \ldots, y_n)$
are tuples of variables of sort $(\sigma_1, \ldots, \sigma_n)$ and
$\varphi_{pre}$, $\varphi_{trans}$ and $\varphi_{post}$ are formulas.

When this command is well-formed, given the above definitions,
this command is syntax sugar for:
\begin{lstlisting}[language=SyGuS-Desugar]
($\vardeclkwd$ $v_1$ $\sigma_1$)
($\vardeclkwd$ $v'_1$ $\sigma_1$)
$\,\ldots$
($\vardeclkwd$ $v_n$ $\sigma_n$)
($\vardeclkwd$ $v'_n$ $\sigma_n$)
($\constraintkwd$ (=> ($S_{pre}$ $v_1$ $\ldots$ $v_n$) ($S$ $v_1$ $\ldots$ $v_n$)))
($\constraintkwd$ (=> (and ($S$ $v_1$ $\ldots$ $v_n$) ($S_{trans}$ $v_1$ $\ldots$ $v_n$ $v'_1$ $\ldots$ $v'_n$)) ($S$ $v'_1$ $\ldots$ $v'_n$)))
($\constraintkwd$ (=> ($S$ $v_1$ $\ldots$ $v_n$) ($S_{post}$ $v_1$ $\ldots$ $v_n$)))
\end{lstlisting}
where $v_1, v'_1, \ldots, v_n, v'_n$ are fresh symbols.

\item \code|($\constraintchckwd$ (($x_1$ $\sigma_1$) $\ldots$ ($x_m$ $\sigma_m$)) $t_{body}$ $t_{head}$)|

This command adds a universally quantified implication constraint to the current state
in the form of a constrained Horn clause (CHC).
Concretely, it adds $m$ fresh variables $(v_1, \ldots, v_m)$ of sorts $(\sigma_1, \ldots, \sigma_m)$ respectively
to the list of universal variables,
and a constraint
$$
  ((\lambda x_1,\ldots,x_m \ldotp t_{body})\ v_1\ \ldots\ v_m) \implies ((\lambda x_1,\ldots,x_m \ldotp t_{head})\ v_1\ \ldots\ v_m)
$$
to the current state.
This is a generalization of the $\constraintinvkwd$ command,
that allows for specifying problems with multiple, possibly interdependent,
invariant predicates to be synthesized.

A constraint of this form is well-formed if
the \emph{body} of the CHC (i.e., $t_{body}$) and the \emph{head} of the CHC (i.e., $t_{head}$)
are both well-sorted terms of $\sbool$ sort.

When this command is well-formed, given the above definitions,
this command is syntax sugar for:
\begin{lstlisting}[language=SyGuS-Desugar]
($\vardeclkwd$ $v_1$ $\sigma_1$)
$\,\ldots$
($\vardeclkwd$ $v_m$ $\sigma_m$)
($\fundefkwd$ $F_{body}$ (($x_1$ $\sigma_1$) $\ldots$ ($x_m$ $\sigma_m$)) Bool $t_{body}$)
($\fundefkwd$ $F_{head}$ (($x_1$ $\sigma_1$) $\ldots$ ($x_m$ $\sigma_m$)) Bool $t_{head}$)
($\constraintkwd$ (=> ($F_{body}$ $v_1$ $\ldots$ $v_m$) ($F_{head}$ $v_1$ $\ldots$ $v_m$)))
\end{lstlisting}
where $v_1, \ldots, v_m$, $F_{body}$, and $F_{head}$ are fresh symbols.

\end{itemize}

\subsection{Asserting Oracle Constraints and Assumptions}
\begin{itemize}
\item $\paren{\constraintoraclekwd\mbox{ }  
\paren{\paren{x_1\mbox{ }\sigma_1} \ldots \paren{x_n\mbox{ }\sigma_n}}\mbox{ }
\paren{\paren{y_{1}\mbox{ }\tau_{1}} \ldots \paren{y_{m}\mbox{ }\tau_{m}}}  \mbox{ } t \mbox{ } N }$

This command informs the solver of the existance of an external binary with name $N$
which can be used as means of adding new constraints to the problem.
This command is well-formed only if $N$
\emph{implements} a function of sort $\sigma_1 \times \ldots \times \sigma_n \rightarrow \tau_{1} \times \ldots \times \tau_{m}$
(for a definition of the expected implementation of an external binary, see \cref{sec:oracleimplementations}),
and $t$ is a well-sorted term of sort $\sbool$ 
whose free symbols may include those in the current signature, as well as any symbol in $x_1 \ldots x_{n}$ and $y_1 \ldots y_{m}$,
and moreover is allowed based on the restrictions of the current logic.

Assuming this command is well-formed,
the interaction between the synthesis solver and the binary
can be understood as the solver passing values
$c_1, \ldots, c_n$ for $x_1, \ldots x_n$ as input to the binary,
and the binary generating a list of values
$d_{1}, \ldots, d_{m}$ corresponding to the output $y_{1}, \ldots y_{m}$.
The expected implementation for passing
constant values as input and output 
is described in \cref{sec:oracleimplementations}.
For each such call,
the formula $t[c_1 \ldots c_{n} d_1 \ldots d_m]$, i.e., 
$t$ with all occurences of $x_1, \ldots, x_{n}, y_1, \ldots, y_m$ replaced with $c_1, \ldots, c_{n}, d_1, \ldots d_m$, 
is added to the current list of constraints $\varphi$ in the conjecture.

Note that 
the synthesis solver may choose to call the binary $N$ 
any time during solving, and as many times as it chooses.
Moreover, the solver may call the binary with
the same input more than once.

\item 
$\paren{\assumeoraclekwd\mbox{ }
\paren{\paren{x_1\mbox{ }\sigma_1} \ldots \paren{x_n\mbox{ }\sigma_n}}\mbox{ }
\paren{\paren{y_{1}\mbox{ }\tau_{1}} \ldots \paren{y_{m}\mbox{ }\tau_{m}}}  \mbox{ } t \mbox{ } N }$

This command is identical to $\constraintoraclekwd$,
but the term $t[c_1 \ldots c_{n} d_1 \ldots d_m]$ obtained from a call to the external binary
is added to the set of assumptions instead of the set of constraints.

\end{itemize}

\subsection{Declaring Oracle Functional Symbols}

\begin{itemize}
\item $\paren{\oracledeckwd\mbox{ }S\mbox{ } \paren{\sigma_1 \ldots \sigma_n}\mbox{ }\sigma \mbox{ } N}$

This adds to the current signature a symbol $S$ of function sort 
$\sigma_1 \times \ldots \times \sigma_n \rightarrow \sigma$
whose interpretation is given by an external oracle also with name $N$. 
For every call to the oracle, an assumption about the behavior of $S$ is added to the list of assumptions $\alpha$ in the conjecture.
Note that this command is syntactic sugar for:

\begin{lstlisting}[language=SyGuS-Desugar]
($\vardeclkwd$ $S$ (-> $\sigma_1$ $\ldots$ $\sigma_n$ $\sigma$))
($\assumeoraclekwd$ (($x_1$ $\sigma_1$) $\ldots$ ($x_n$ $\sigma_n$)) (($x$ $\sigma$)) (= ($S$ $x_1$ $\ldots$ $x_n$)) $N$)
\end{lstlisting}

where $x_1, \ldots, x_n$ and $x$ are fresh variables.
\end{itemize}

\subsection{Pre-defined Oracle Types}
Given a synthesis function symbol $F$ in the current signature
with sort $\sigma_1, \ldots \sigma_n \rightarrow \sigma$, each of the following commands 
is used to declare an external oracle with name $N$, which generate specific types of oracle constraints and assumptions over the behavior of $F$. 

\begin{itemize}
\item $\paren{\iooracledeckwd\mbox{ }F\mbox{ } N}$

This command declares an input-output oracle for function $F$. 
It is syntactic sugar for:

\begin{lstlisting}[language=SyGuS-Desugar]
($\constraintoraclekwd$
  (($x_1$ $\sigma_1$) $\ldots$ ($x_n$ $\sigma_n$))
  (($x$ $\sigma$))
  (= ($F$ $x_1$ $\ldots$ $x_n$) $x$)
  $N$)
\end{lstlisting}

\item$\paren{\poswitnessoracledeckwd\mbox{ }F\mbox{ } N} $

This command declares a positive witness oracle for synthesis function $F$.
It is syntactic sugar for:

\begin{lstlisting}[language=SyGuS-Desugar]
($\constraintoraclekwd$
  ()
  (($x_1$ $\sigma_1$) $\ldots$ ($x_n$ $\sigma_n$) ($x$ $\sigma$))
  (= ($F$ $x_1$ $\ldots$ $x_n$) $x$)
  $N$)
\end{lstlisting}

\item$\paren{\negwitnessoracledeckwd\mbox{ }F\mbox{ } N}$

This command declares a negative witness oracle for synthesis function $F$.
It is syntactic sugar for:

\begin{lstlisting}[language=SyGuS-Desugar]
($\constraintoraclekwd$
  ()
  (($x_1$ $\sigma_1$) $\ldots$ ($x_n$ $\sigma_n$) ($x$ $\sigma$))
  (not (= ($F$ $x_1$ $\ldots$ $x_n$) $x$))
  $N$)
\end{lstlisting}

\item$\paren{\memoracledeckwd\mbox{ }F\mbox{ } N} $

This command declares a membership-query oracle for synthesis function $F$.
It is syntactic sugar for:

\begin{lstlisting}[language=SyGuS-Desugar]
($\constraintoraclekwd$
  (($x_1$ $\sigma_1$) $\ldots$ ($x_n$ $\sigma_n$) ($x$ $\sigma$))
  (($R$ Bool))
  (= (= ($F$ $x_1$ $\ldots$ $x_n$) $x$) $R$)
  $N$)
\end{lstlisting}

\item$\paren{\cexoracledeckwd\mbox{ }F\mbox{ } N} $

This command declares a counter-example oracle for synthesis function $F$.
It is syntactic sugar for:

\begin{lstlisting}[language=SyGuS-Desugar]
($\constraintoraclekwd$
  (($F_c$ (-> $\sigma_1$ $\ldots$ $\sigma_n$ $\sigma$)))
  (($R$ Bool) ($x_1$ $\sigma_1$) $\ldots$ ($x_n$ $\sigma_n$))
  (=> $R$ (not (= ($F$ $x_1$ $\ldots$ $x_n$) ($F_c$ $x_1$ $\ldots$ $x_n$))))
  $N$)
\end{lstlisting}

where $F_c$ is a candidate implementation for $F$, and $R$ is a boolean that indicates that the oracle was able to find a counterexample.

\item$\paren{\corroracledeckwd\mbox{ }F\mbox{ } N} $

This command declares a correctness oracle that determines
whether a candidate implementation of $F$ is correct.
It is syntactic sugar for:

\begin{lstlisting}[language=SyGuS-Desugar]
($\oracledeckwd$ $s$ ((-> $\sigma_1$ $\ldots$ $\sigma_n$ $\sigma$)) Bool $N$)
($\constraintkwd$ ($s$ $F$))
\end{lstlisting}

where $s$ is a fresh symbol. 

\item$\paren{\corrcexoracledeckwd\mbox{ } F\mbox{ }N} $

This command declares a correctness oracle that determines
whether a candidate implementation of $F$ is correct,
and returns a counterexample if the implementation is found to be incorrect.
It adds a new constraint to the set of oracle constraints.
It is syntactic sugar for:

\begin{lstlisting}[language=SyGuS-Desugar]
($\constraintoraclekwd$
  (($F_c$ (-> $\sigma_1$ $\ldots$ $\sigma_n$ $\sigma$)))
  (($R$ Bool) ($x_1$ $\sigma_1$) $\ldots$ ($x_n$ $\sigma_n$))
  (=> $R$ (not (= ($F$ $x_1$ $\ldots$ $x_n$) ($F_c$ $x_1$ $\ldots$ $x_n$))))
  $N$)
($\oracledeckwd$ $s$ ((-> $\sigma_1$ $\ldots$ $\sigma_n$ $\sigma$)) Bool $N$)
($\constraintkwd$ ($s$ $F$))
\end{lstlisting}

where $s$ is a fresh symbol.


\end{itemize}

\subsection{Initiating Synthesis Solver}

\begin{itemize}
\item $\paren{\checksynthkwd}$

This command asks the synthesis solver to find a solution for the synthesis conjecture
corresponding to the current list of functions-to-synthesize,
universal variables and constraints.

\item $\paren{\optsynthkwd\mbox{ }\paren{s_1 \ldots s_j}\mbox{ } A_1 \ldots A_k}$

Like the command above,
this asks the synthesis solver to find a solution for the synthesis conjecture
corresponding to the current list of functions-to-synthesize,
universal variables and constraints.
Moreover, that solution should be optimal with respect to the objective
specified by the list of terms $s_1, \ldots, s_j$ and attributes $A_1, \ldots, A_k$.
Objectives can be specified both using annotations on terms $s_1, \ldots, s_j$
as well as the use of attributes $A_1, \ldots, A_k$ to specify orderings on tuples.
Details on objectives in given in \cref{sec:attr-objectives}.

To be well formed, $s_1, \ldots, s_j$ cannot contain any occurrence
of universal variables in the current state,
and must be permitted by the background logic.
\end{itemize}

The expected output
from the synthesis solver for these commands is covered in \cref{sec:output}.

\subsection{Setting Benchmark Information}

\begin{itemize}
\item $\paren{\setinfokwd\mbox{ }:S\mbox{ }L}$

This command sets meta-information specified by the symbol $S$
to the (literal) value $L$, whose syntax is given in \cref{ssec:literals}.
This has no impact on the state, and is used to annotate the
benchmark with relevant information.
For the purposes of this document, we define the meaning of
the concrete symbol, ${\tt sygus\mbox{-}version}$,
which when passed to this command is used to indicate the version of the SyGuS
format used in the benchmark.
A benchmark whose header contains the line
\[
\paren{\setinfokwd\mbox{ }:{\tt sygus\mbox{-}version\mbox{ }2.1}}
\]
will use the version 2.1 syntax as specified by this document,
whereas e.g. a value of $1.0$ indicates that the benchmark will use the
syntax in the previous SyGuS format~\cite{sygusFormat,sygusSyntax2015,sygusSyntax2016}.
\end{itemize}
\subsection{Setting Solver Options}

\begin{itemize}
\item $\paren{\setoptkwd\mbox{ }:S\mbox{ }L}$

This command sets the solver-specific option specified by the symbol $S$ 
to the (literal) value $L$, whose syntax is given in \cref{ssec:literals}.
We do not give concrete examples of such options in this document. 
It is recommended that synthesis solvers
ignore unrecognized options, 
and choose reasonable defaults when the
options are left unspecified.
\end{itemize}

\section{Synthesis Solver Output}%
\label{sec:output}

This section covers the expected output from a synthesis solver,
which currently is limited to responses to $\checksynthkwd$ 
only.

\begin{itemize}
\item
A response to $\checksynthkwd$ is one of the following:

\begin{enumerate}
\item
A list of commands, enclosed in parentheses, of the form:
\[
\begin{array}{l}
\paren{\\
\paren{\fundefcmdkwd\mbox{ }f_{1}\mbox{ }X_{1}\mbox{ }\sigma_{1}\mbox{ }t_{1}}\\
\ldots\\
\paren{\fundefcmdkwd\mbox{ }f_{n}\mbox{ }X_{n}\mbox{ }\sigma_{n}\mbox{ }t_{n}}\\
}
\end{array}
\]
where functions $f_{1}, \ldots, f_{n}$
are the functions-to-synthesize in the current state,
$X_{1}, \ldots, X_{n}$ are sorted variable lists,
$\sigma_{1}, \ldots, \sigma_{n}$ are types,
and $t_{1}, \ldots, t_{n}$ are terms.
The syntax $\fundefcmdkwd$ can be either $\fundefkwd$ or $\recfundefkwd$.
The latter must be used for $f_i$ if it occurs free in $t_i$,
that is, when the definition of $f_i$ is recursive.
It is required that $f_1, \ldots, f_n$ be
provided in the order in which they were declared.\footnote{
This is to ensure that the definition of $f_i$, which may depend on a definition
of $f_j$ for $j<i$, is given in the correct order.
}
To be a well-formed response, 
for each $j=1, \ldots, n$, it must be the case that
$t_{j}$ is a term of $\sigma_{j}$,
and $X_{j}$ is identical to the sorted variable list
used when introducing the function-to-synthesize $f_{j}$.

\item
The output ${\tt infeasible}$,
indicating that the conjecture has no solutions.

\item
The output ${\tt fail}$,
indicating that the solver failed to find a solution to the conjecture.

\end{enumerate}

A response to $\checksynthkwd$ of the first kind is a correct solution if
it satisfies both the semantic and syntactic restrictions given by
the current state.
We describe this in more detail in \cref{sec:logical-semantics}.

We do not define the correctness of an ${\tt infeasible}$ response 
in this document. 
We remark that a response of this form should be given
by a solver only if it is certain that a solution to the current
conjecture does not exist.
A conjecture may be infeasible based on the current semantic
restrictions, or may be infeasible due to a combination
of semantic restrictions and the syntactic ones imposed on functions-to-synthesize.
The response ${\tt fail}$ indicates that the solver
was unable to find a solution, which does not necessarily imply that
the conjecture is infeasible.

\item
A (successful) well-formed response to $\paren{\optsynthkwd\mbox{ }\paren{s_1 \ldots s_j}\mbox{ } A_1 \ldots A_k}$
is a tuple of terms followed list of commands, enclosed in parentheses, of the form:
\[
\begin{array}{l}
\paren{\\
\paren{c_1 \ldots c_j}\\
\paren{\fundefcmdkwd\mbox{ }f_{1}\mbox{ }X_{1}\mbox{ }\sigma_{1}\mbox{ }t_{1}}\\
\ldots\\
\paren{\fundefcmdkwd\mbox{ }f_{n}\mbox{ }X_{n}\mbox{ }\sigma_{n}\mbox{ }t_{n}}\\
}
\end{array}
\]
To be a well-formed response, 
$c_1, \ldots, c_j$ must be values of the same type as $s_1, \ldots, s_j$,
and the remaining commands must be a well-formed response based on the criteria
for $\checksynthkwd$ above.
A response of the above form is correct if
the specified functions satisfy the syntactic restrictions given by
the current state.
Additionally, the solution must satisfy
the semantic specification for optimization queries, which
implies that both the semantic specification is satisfied
and that the values $c_1, \ldots, c_j$
are consistent valuations of terms $s_1, \ldots, s_j$
under the current solution for functions to synthesize $f_1, \ldots, f_n$.
For details, see \cref{sec:logical-semantics}.

Additionally, analogous to $\checksynthkwd$, 
a solver may respond
with ${\tt infeasible}$ or ${\tt fail}$ to indicate
the conjecture has no solutions, or to indicate a failure.

The role of $A_1 \ldots A_k$ is to provide an ordering on solutions
generated in response to this command, where a solver is expected to generate
a solution that is optimal with respect to this ordering.
For details, see \cref{sec:attr-objectives}.
These attributes do not impact what consistutes a correct solution.

\end{itemize}


\section{SyGuS Background Logics}%
\label{sec:sygus-logic}

In this section,
we describe how the background logic
restricts the constraints and grammars that are allowed as inputs.
A SyGuS background logic consists of three parts:
\begin{enumerate}
\item An \emph{input logic}, which can be set as part of a $\setlogickwd$ command.
This corresponds to a SMT-LIB standard logic or may be solver specific.
The input logic determines the set of terms that are allowed in 
constraints.

\item An \emph{output logic}, which can be set as part of a $\setlogickwd$ command.
Like input logics, this
can correspond to an SMT-LIB standard logic or may be solver specific.
The output logic determines the set of terms that are allowed in
grammars and solutions.

\item A \emph{feature set}, 
which restricts or extends the set of commands that are allowed in a SyGuS input,
and may further refine the constraints, grammars and solutions allowed by the logic.
Generally, feature sets are not expressible in an input or output logic
and are independent of them.
\end{enumerate}
We refer to input and output logics as \emph{base} logics.
Base logics have the same scope as SMT-LIB logics, in that their purpose is to
define a set of terms and formulas.
Further extensions and restrictions of the SyGuS language are recommended
to be expressed as new base logics whenever it is possible to do so.
On the other hand,
the feature set is allowed to restrict or extend the
\emph{commands} that are allowed in the input
or specific relationships between how terms appear in commands,
which is not expressible in a base logic.

The default SyGuS logic is the one
whose input and output logics include only the core theory of Booleans,
and whose feature set enables grammars and the core commands of the language.
We describe logics and feature sets
in more detail in \cref{ssec:smt-logic,ssec:feature-sets}.
A formal definition of how these
restrict the set of terms that may appear 
in constraints and grammars
is then given in \cref{ssec:logicr-constraints,ssec:logicr-grammars}.

\subsection{Input and Output Logics}
\label{ssec:smt-logic}

SMT-LIB provides a catalog of standard logics,
available at \url{www.smt-lib.org}.
SyGuS logics may use any of these logics
either as input logics or as output logics.
For many applications,
the input and output logic is expected to be the same, although
no relationship between the two is required.

At a high level, a \emph{logic} includes a set of theories
and defines a subset of terms constructible in the signature of those theories
that belong to it.
If a logic includes a theory, then its symbols are added
to the current signature when a $\setlogickwd$ command is issued.
Further details on the formal definition of 
theory and logic declarations can be found in Sections 3.7 and 3.8
of the SMT-LIB 2.6 standard~\cite{BarFT-RR-17}.
We briefly review some of the important logics and theories in the following.
More concrete details on
standard SMT-LIB logics can be found in the reference
grammars in \cref{apx:ref-grammars}.

The input and output logic components of the default SyGuS logic include only the
\emph{core theory}.
The signature of the core theory 
has the Boolean sort ${\tt Bool}$,
the Boolean constants ${\tt true}$ and ${\tt false}$, the
usual logical Boolean connectives 
${\tt not}$, ${\tt and}$, ${\tt or}$, implication ${\tt =>}$, ${\tt xor}$,
and the parametric symbols ${\tt =}$ and ${\tt ite}$
denoting equality and if-then-else terms for all sorts in the signature.
Logics that include the theories mentioned in this section supplement the signature
of the current state with additional sorts and symbols of that theory.

\paragraph{Arithmetic}
The theory of integers is enabled 
in logics like linear integer arithmetic ${\tt LIA}$
or non-linear integer arithmetic ${\tt NIA}$.
The signature of this theory includes the integer sort ${\tt Int}$ and
typical function symbols of arithmetic, including
addition ${\tt +}$
and multiplication ${\tt *}$.
Unary negation and subtraction are specified by ${\tt -}$.
Constants of the theory are integer constants.
Positive integers and zero are specified by the syntax $\intconst$ from \cref{ssec:literals},
whereas negative integers are specified as the unary negation of positive integer,
that is, ${\tt (-\ 2)}$ denotes negative two.
Analogously, 
the theory of reals is enabled
in logics like linear real arithmetic ${\tt LRA}$
or non-linear real arithmetic ${\tt NRA}$.
Its signature includes the real sort ${\tt Real}$.
Some of the function symbols of arithmetic are syntactically identical to
those from the theory of integers, including
${\tt +}$, ${\tt -}$ and ${\tt *}$.
The signature of the theory of reals additionally includes
real division ${\tt /}$.
Positive reals and zero in this theory can either be specified as
decimals using the syntax $\realconst$
or as rationals of the form ${\tt (/\ m\ n)}$,
where ${\tt m}$ and ${\tt n}$ are numerals.
Negative reals are specified as the unary negation of a decimal
or as a negative rational ${\tt (/\ (-\ m)\ n)}$.

\paragraph{Bit-Vectors}
The theory of fixed-width bit-vectors
is included in logics specified 
by symbols that include the substring ${\tt BV}$.
The signature of this theory includes a family of 
indexed sorts ${\tt (\_\ BitVec\ n)}$
denoting bit-vectors of width $n$.
The functions in this signature include various operations on bit-vectors,
including bit-wise, arithmetic, and shifting operations.

\paragraph{Strings}
The theory of (unbounded) Unicode strings and regular expressions
is included in some logics specified with ${\tt S}$ as a prefix,
such as ${\tt S}$ (strings) or ${\tt SLIA}$ (strings
with linear integer arithmetic).
The string of this theory includes the string sort ${\tt String}$,
interpreted as the set of all unicode strings.
Functions in this signature include string concatenation ${\tt str.++}$,
string length ${\tt str.len}$ as well as many extended functions
such as string containment ${\tt str.contains}$,
string search ${\tt str.indexof}$ which returns the index of
a string in another, and so on.
A full description of this theory is given in a proposal to SMT-LIB~\cite{smtlibStrings}.

\paragraph{Arrays}
The theory of arrays is included in logics specified with ${\tt A}$ as a prefix,
such as ${\tt ABV}$ or ${\tt ALIA}$.
The signature of this theory includes a parametric sort ${\tt Array}$
of arity two, whose sort parameters
indicate the index type and the element type of the array.
It has two function symbols, ${\tt select}$ and ${\tt store}$,
interpreted as array select and array store.

\paragraph{Datatypes}
The theory of datatypes
is included in logics specified 
by symbols that include the substring ${\tt DT}$.
Logics that include datatypes are such that
$\dtsdeclkwd$ commands are permitted in inputs, 
whereas all others do not.
The signature of the theory of datatypes
is largely determined by the concrete datatypes definitions provided
by the user.
As mentioned in \cref{ssec:declaring-sorts}, these commands append datatype sorts, 
constructors and selectors to the current signature.
Constructor symbols are used for constructing
values (e.g. ${\tt cons}$ constructs a list from an element and another list),
and selectors access subfields (e.g. ${\tt tail}$ returns its second argument).
Notice that the value of \emph{wrongly applied} selectors, e.g. ${\tt tail}$ applied
to the ${\tt nil}$ list, is underspecified and hence freely interpreted
in models of this theory.
The only fixed symbol in the theory of datatypes is the unary
indexed \emph{discriminator} predicate ${\tt(\_\ is\ C)}$,
which holds if and only if its argument is an application of constructor $C$.
For example, assuming the standard definition of a list datatype
with constructors ${\tt cons}$ and ${\tt nil}$,
we have that ${\tt ((\_\ is\ nil)\ x)}$ holds if and only if
$x$ is the ${\tt nil}$ list.

\paragraph{Uninterpreted Functions}
In SMT-LIB, uninterpreted functions and sorts may be declared in
logics that include the substring ${\tt UF}$,
whose interpretations are not fixed.
Declarations for functions and sorts are made via SMT-LIB commands 
$\fundeclkwd$ and $\sortdeclkwd$ respectively.
In the SyGuS language, we do not permit
the declaration of functions with $\fundeclkwd$ command.
Instead, the language includes only the latter command.
Thus, the only effect that specifying ${\tt UF}$ in the logic string has is
that user-defined sorts may be declared via $\sortdeclkwd$, where
variables and functions-to-synthesize may involve these sorts in the usual way.
We remark here that
encoding synthesis problems 
that involve (existentially quantified)
uninterpreted functions
can be represented by declaring those functions
using $\synthfunkwd$ commands where no grammar is provided.
Synthesis problems that involve universally quantified variables of function sort
are planned to be addressed in a future revision of this document
that includes concrete syntax for function sorts.

\subsubsection{Other Theories}

Many other theories
are possible beyond those supported in the SMT-LIB standard.
In this section, we mention theories that are of interest to synthesis applications
that are not included in the SMT-LIB standard.

\paragraph{Bags and Tables}
An SMT-LIB compliant theory of tables is proposed in~\cite{theoryOfTables}.
We give the salient details of this theory here.

The theory of tables is implemented as an extension of a theory of bags (multisets).
The signature of this theory includes all sorts of the form
$\texttt{(Bag\ T)}$
for all sorts $\texttt{T}$. 
This sort denotes bags (i.e. multisets) of elements of sort $\texttt{T}$.
The theory of bags includes a multitude of operators for e.g.
taking unions, intersections and differences of bags,
as well as higher-order operators like 
$\texttt{map}$, $\texttt{fold}$ and $\texttt{partition}$.
A \emph{table} is a bag whose element sort $T$ is a tuple,
where a tuple is a parametric sort taking $n$ types corresponding to the
types of the elements that comprise the tuple.
For example, $\texttt{(Tuple\ Int\ Int)}$ denotes tuples
of arity two.
The sort $\texttt{(Table\ Int\ Int)}$ is syntax sugar for
$\texttt{(Bag\ (Tuple\ Int\ Int))}$ and is used to denote tables with two integer
columns.
A table value is a multiset union of tuple values.
The number of rows in a table value is equal to its cardinality.
Notice that ordering of rows is not semantically captured,
and thus not modelled in this theory.
For further details on the complete signature and semantics of this
theory is available in~\cite{theoryOfTables}.

\subsection{Features and Feature Sets}
\label{ssec:feature-sets}

A feature set is a set of values, called \emph{features},
which for the purposes of this document can be seen as an enumeration type.
Their syntax is given in \cref{ssec:syntax-features}.
The meaning of all features standardized by this document are listed below.
\begin{itemize}
\item $\fgrammarskwd$: if enabled, 
then grammars may be provided for functions-to-synthesize 
in $\synthfunkwd$ commands.
\item $\ffwddeclskwd$: if enabled,
grammars of $\synthfunkwd$ may refer to previously declared synthesis functions,
called \emph{forward declarations}.
\item $\frecursionkwd$: if enabled,
grammars of $\synthfunkwd$ can generate terms that correspond to recursive definitions.
\item $\foracleskwd$: if enabled,
commands from $\oraclecmd$ are permitted.
\item $\fweightskwd$: if enabled,
terms of the form \weightsym{w}{f} are permitted,
as described below.
\end{itemize}
Formal definition of these features are given within the 
\cref{ssec:logicr-constraints,ssec:logicr-grammars}.
Other features and their meanings may be solver specific, 
which are not covered here.

The feature set component of the default SyGuS logic is the set $\{ \fgrammarskwd \}$.
In other words, grammars may be provided, but those involving
forward declarations and recursion are not permitted by default,
nor are oracles or weight constraints.

\subsubsection{Weights}
\label{sec:weight-semantics}
For each function-to-synthesize $f$ and weight keyword $W$
as mentioned in \cref{sec:weight-attributes},
the \emph{weight symbol for $f$ with respect to $W$} is a (nullary) integer symbol $\wsym{f}{W}$
whose concrete syntax is \weightsymnf{$w$}{$f$} where $w$ is the
suffix of $W$ after its first character ($:$).
For example, \weightsym{weight}{f}
denotes a symbolic integer constant corresponding to the weight of $f$
with respect to the builtin weight keyword.
We will use annotations on terms that appear in grammars
to denote weighted production rules.
The class of nullary integer symbols of the above form are
interpreted as the sum of weights of production rules used for generating the body of $f$.

Formally,
let $G$ be a grammar provided for function-to-synthesize $f$, and let $W$ be a weight keyword.
Let $y$ be a non-terminal symbol of $G$ and let $t$ be a term appearing
in the grouped rule listing for $y$, as described in the syntax for grammar
definitions in \cref{ssec:dec-synth-fun}.
We consider the \emph{base term} of $t$, defined such that
the base term of unannotated term is itself, and
the base term of $t$ is $t_0$ if $t$ is of the form 
$\paren{!\mbox{ }t_0\mbox{ }A_1\mbox{ }\ldots\mbox{ }A_n}$.
Notice that nested annotations on $t_0$ are not considered in this defintion.
Assume that the base term of $t$ is $t_0$ and that
$t$ has been annotated with the (possibly empty) list of attributes $A_1, \ldots, A_n$.
In this case, we say that $G$ contains
the production rule $y \mapsto t_0$ whose \emph{weight with respect to $W$} is:
\[
\begin{cases}
k & \text{if exactly one } A_i \text{ is of the form } \texttt{W}\ k \text{ for some numeral } k \\
k_{def} & \text{otherwise, where } k_{def} \text{ is the default weight value for } \texttt{W}
\end{cases}
\]
The default weight value for all weight keywords is $0$.
This value can be overridden via an attribute as described in \cref{sec:attr-weights}.

Let $y \mapsto_{W,k} t_0$ denote a production rule
whose weight with respect to $W$ is $k$.
We write $\wgenerates{G}{W}{k}{r}$ if it is possible to construct a sequence of terms
$s_1, \ldots, s_n$
with $s_1$ is the starting symbol of $G$ and $s_n = r$
where for each $1 \leq i \leq n$, term $s_i$ is obtained from $s_{i-1}$ by
replacing an occurrence of some $y_i$ by $t_i$
where $y_i \mapsto_{W,k_i} t_i$ is a rule in $G$,
and $k_1 + \ldots k_n = k$.
When the body of function-to-synthesize $f$ is interpreted as term $r$,
the symbol $\wsym{f}{W}$ can be interpreted as $k$ if and only if
$\wgenerates{G}{W}{k}{r}$.

Notice that there may be \emph{multiple} sequences of terms that
generate the same term $r$ above.
Thus, weight symbols may have multiple valid
interpretations for a given interpretation for the body of a function-to-synthesize.
The solver is free to pick any such interpretation.

\subsection{SyGuS Logic Restrictions on Constraints}
\label{ssec:logicr-constraints}


Let $\slogic$ be a SyGuS logic whose input logic
is one from the SMT-LIB standard, or an externally defined logic.
A term $t$ is not allowed by $\slogic$ 
to be an argument to a $\constraintkwd$ command 
if it is not allowed by the \emph{input} logic of $\slogic$, 
according to the definition of that logic.



\subsection{SyGuS Logic Restrictions on Grammars}
\label{ssec:logicr-grammars}

Let $\slogic$ be a SyGuS logic whose output logic
is one from the SMT-LIB standard.
A grammar $G$ is not allowed by $\slogic$ if
it generates some term $t$ with 
no free occurrences of non-terminal
symbols that is not allowed by the \emph{output} logic of $\slogic$,
according to the definition of that logic. 

Notice that it may be the case that a grammar $G$
contains a rule whose conclusion is a term 
that does not itself meet the restrictions 
of the output logic.
For example, consider the logic of \emph{linear} integer arithmetic
and a grammar $G$ containing a non-terminal symbol $y_c$ of integer type
such that $G$ generates only constants from $y_c$.
Grammar $G$ may be allowed in the logic of linear integer arithmetic
even if it has a rule whose conclusion is $y_c * t$, 
noting that no non-linear terms can be generated from this rule,
provided that no non-linear terms can be generated from $t$.
An example demonstrating this case is given in \cref{sec:examples}.

The feature set component of the SyGuS logic 
imposes additional restrictions
on the terms that are generated by grammars.
Note the following definition.
The \emph{expanded form} of a term $t$
is the (unique) term obtained
by replacing all functions $f$ in $t$ that are defined as macros
with their corresponding definition until a fixed point is reached.
A grammar $G$ for function-to-synthesize $f$ is not allowed by a SyGuS logic $\slogic$
if $G$  contains a rule whose conclusion is a term $t$
whose expanded form contains applications of functions-to-synthesize
unless the feature $\ffwddeclskwd$ is enabled in the feature set of $\slogic$;
it is not allowed if $t$ 
contains $f$ itself unless the feature $\frecursionkwd$ is enabled in the feature set of $\slogic$;
it is not allowed regardless of the terms it generates unless $\fgrammarskwd$ is enabled in the feature set of $\slogic$.

\subsection{Additional SyGuS Logics}
Here, we cover additional SyGuS logics that are standardized
by this document that are \emph{not} defined by the SMT-LIB standard.

\paragraph{Programming-by-examples (PBE)}
Given an SMT-LIB standard logic $X$ 
that does not contain the prefix ${\tt QF\_}$,
the base logic ${\tt PBE\_}X$ denotes the logic
where constraints are limited to (conjunctions of)
equalities whose left hand side is a term $f( \vec c )$ and 
whose right hand side is $d$,
where $f$ is a function
and $\vec c$, $d$ are constants.
We refer to an equality of this form as a \emph{PBE equality}.
Such equalities denote a relationship between the inputs and output of 
function $f$ for a single point.
Notice that formulas allowed by 
logic ${\tt PBE\_}X$ are a subset of those allowed by ${\tt QF\_}X$.

We use the logic string ${\tt PBE\_}X$ to refer to a SyGuS logic as well.
Given an SMT-LIB standard logic $X$
that does not contain the prefix ${\tt QF\_}$,
the SyGuS logic ${\tt PBE\_}X$ is the one
whose input logic is ${\tt PBE\_}X$,
whose output logic ${\tt QF\_}X$,
and whose feature set is the default one.
In other words, the constraints allowed by this logic
are limited to conjunctions of PBE equalities,
but gives no special restrictions
on the solutions or grammars that can be provided.

By construction of the overall synthesis conjecture, a SyGuS command sequence
meets the requirements of the input logic ${\tt PBE\_}X$
if and only if each ${\tt constraint}$ command takes as argument a PBE equality.

\paragraph{Single Invariant-to-Synthesize (Inv)}
Given an SMT-LIB standard logic $X$,
the base logic \code|Inv_$X$| denotes the logic
where formulas are limited to the invariant synthesis
problem for a single invariant-to-synthesize.
Concretely,
this means that formulas are limited to those that 
are (syntactically) a conjunction of three implications, 
for a single predicate symbol $I$ to be synthesized:
\begin{enumerate}
\item The first is an implication whose antecedent (the pre-condition) 
is an arbitrary formula
in the logic $X$ and
and whose conclusion is
an application $I( \vec x )$ where $\vec x$ is a tuple of unique variables,

\item The second is an implication whose antecedent is
$I( \vec x ) \wedge \varphi$
where $\vec x$ is a tuple of unique variables and $\varphi$ (the transition relation) 
is an arbitrary formula
in the logic $X$,
and whose conclusion is $I( \vec y )$
where $\vec y$ is a tuple of unique variables disjoint from $\vec x$,

\item The third is an implication whose antecedent is 
an application $I( \vec x )$ where $\vec x$ is a tuple of unique variables,
and whose conclusion (the post-condition) is an arbitrary formula in the logic $X$.
\end{enumerate}
The variables $\vec x$ in each of these three formulas are not required to be the same.

Like the previous section, 
we use the logic string \code|Inv_$X$| to refer to SyGuS logics as well.
Given an SMT-LIB standard logic $X$
that does not contain the prefix ${\tt QF\_}$,
logic \code|Inv_$X$| denotes the SyGuS logic
whose input logic is \code|Inv_QF_$X$|,
whose output logic is \code|QF_$X$|,
and whose feature set is the default one.

Note that if a SyGuS command sequence is such that it contains
(1) exactly one $\synthfunkwd$ command with $\boolkwd$ return type,
(2) exactly one $\constraintinvkwd$ command whose arguments are predicates with definition within logic $X$, and
(3) no other commands that introduce constraints,
then the command sequence is guaranteed to meet the requirements of the input logic \code|Inv_$X$|.

\paragraph{Constrained Horn Clause Systems (CHC)}
Given an SMT-LIB standard logic $X$,
the base logic \code|CHC_$X$| denotes the logic
where formulas are restricted to a valid system of constrained Horn clauses (CHCs),
i.e., the formulas (syntactically) are a conjunction of CHCs
over a, possibly empty, set of predicate symbols to be synthesized.
To define a valid system of CHCs, we first define the notion of an \emph{atomic term}.

\begin{definition}[Atomic Term]
Given set $\mathbf{S}$ of symbols,
a term $t$ is said to be an $\mathbf{S}$-atomic term if either:
(a) $t$ does not contain any symbol that belongs to $\mathbf{S}$, or
(b) $t$ is of the form \code|($S$ $u_1$ $\ldots$ $u_k$)|
for some predicate symbol $S \in \mathbf{S}$ and $u_1, \ldots, u_k$ are variables.
\qed
\end{definition}

Given this definition, a set $\mathbf{C}$ of CHC constraints
over a set $\mathbf{S}$ of symbols to be synthesized
is considered a valid system of CHCs for logic \code|CHC_$X$|
if $\mathbf{C}$ satisfies the following:
\begin{enumerate}
  \item Each constraint $C \in \mathbf{C}$ satisfies the following:
        \begin{enumerate}
        \item $C$ must be an implication of the form $B_C(\vec x) \implies H_C(\vec x)$,
              where $B_C$ (called the \emph{body} of $C$) and $H_C$ (called the \emph{head} of $C$)
              are defined predicates, and $\vec x$ is a tuple of unique variables.

        \item The definition of $H_C$ must be an $\mathbf{S}$-atomic term.
        
        \item The definition of $B_C$ must either be an $\mathbf{S}$-atomic term,
              or a conjunction of $\mathbf{S}$-atomic terms.
        \end{enumerate}
        Henceforth, we will call such a constraint an $\mathbf{S}$-valid CHC,
        or simply a valid CHC when the set $\mathbf{S}$ is obvious from the context.

  \item Exactly one \emph{query}, i.e., a constraint $C \in \mathbf{C}$ with $H_C$ defined as $\falsekwd$.
\end{enumerate}
The variables $\vec x$ in each of the constraints are not required to be the same.

As in the previous section, 
we use the logic string \code|CHC_$X$| to refer to SyGuS logics as well.
Given an SMT-LIB standard logic $X$
that does not contain the prefix \code|QF_|,
logic \code|CHC_$X$| denotes the SyGuS logic
whose input logic is \code|CHC_QF_$X$|,
whose output logic is \code|QF_$X$|,
and whose feature set is the default one.

A well-sorted SyGuS command sequence meets the requirements of the input logic \code|CHC_$X$|
if it contains
\begin{enumerate}
  \item $\synthfunkwd$ commands with $\boolkwd$ return type only, that declare a set $\mathbf{S}$ of predicates to be synthesized,
  \item one or more $\constraintchckwd$ command each of which is such that
        \begin{enumerate}
          \item its $t_{head}$ is an $\mathbf{S}$-atomic predicate with definition within logic $X$,
          \item its $t_{body}$ is either an $\mathbf{S}$-atomic predicate, or a conjunction of $\mathbf{S}$-atomic predicates,
                each with definitions within logic $X$,
        \end{enumerate}
  \item exactly one $\constraintchckwd$ command with $\falsekwd$ as $t_{head}$, and
  \item no other commands that introduce constraints.
\end{enumerate}

\section{Formal Semantics}%
\label{sec:logical-semantics}

Here we give the formal semantics
for what constitutes a correct solution for a synthesis conjecture.

\subsection{Satisfying Syntactic Specifications}
\label{ssec:sat-syntactic}

In this section,
we formalize the notion of satisfying the \emph{syntactic specification}
of the synthesis conjecture.

As described in \cref{ssec:dec-synth-fun},
a grammar $G$ is specified as a predeclaration and
a \emph{grouped rule listing} of the form
\[
\paren{\paren{y_1\mbox{ }\tau_1\mbox{ }\paren{g_{11} \ldots g_{1m_1} } } \ldots 
\paren{y_n\mbox{ }\tau_n\mbox{ }\paren{g_{n1} \ldots g_{nm_n} } }}
\]
where $y_1, \ldots, y_n$ are variables
and $g_{11} \ldots g_{1m_1}, \ldots, g_{n1} \ldots g_{nm_n}$
are grammar terms.
We associate each grammar with a sorted variable list $X$,
namely the argument list of the function-to-synthesize.
We refer to $y_1$ as the \emph{start symbol} of $G$.

We interpret $G$ as a (possibly infinite) set of rules 
of the form $y \mapsto t$ where $t$ is an (ordinary) term
based on the following definition.
For each $y_i, g_{ik}$ in the grouped rule list,
if $g_{ik}$ is $\paren{\constantkwd\mbox{ }\sigma_c}$,
then $G$ contains the rule $y_i \mapsto c$ for all constants of sort $\sigma_c$.
If $g_{ik}$ is $\paren{\varkwd\mbox{ }\sigma_v}$,
then $G$ contains the rule $y_i \mapsto x$
for all variables $x \in X$ of sort $\sigma_v$.
Otherwise, if $g_{ik}$ is an ordinary term,
then $G$ contains the rule $y_i \mapsto g_{ik}$.

We say that $G$ \emph{generates} term $r$ from $s$
if it is possible to construct a sequence of terms
$s_1, \ldots, s_n$
with $s_1 = s$ and $s_n = r$
where for each $1 \leq i \leq n$, term $s_i$ is obtained from $s_{i-1}$ by
replacing an occurrence of some $y$ by $t$
where $y \mapsto t$ is a rule in $G$.

Let $f$ be a function to synthesize.
A term $\lambda X \ldotp t$
satisfies the syntactic specification for $f$
if one of the following hold:
\begin{enumerate}
\item 
A grammar $G$ is provided for $f$,
$t$ contains no free occurrences of non-terminal symbols,
and $G$ generates $t$ starting from $y_1$,
where $y_1$ is the start symbol of $G$.

\item
No grammar is provided for $f$, and $t$ is any
term allowed in the output logic whose sort
is the same as the return sort of $f$.
\end{enumerate}

Furthermore,
A tuple of functions 
$( \lambda X_1 \ldotp t_1, \ldots, \lambda X_n \ldotp t_n )$
satisfies the syntactic restrictions
for functions-to-synthesize $( f_1, \ldots, f_n )$
in conjecture $\exists f_1, \ldots, f_n \ldotp\Psi$
if $\lambda X_i \ldotp t_i$
satisfies the syntactic specification for $f_i$
for each $i=1,\ldots,n$.

\subsection{Satisfying Semantic Specifications}
\label{ssec:sat-semantic}

In this section,
we formalize the notion of satisfying the \emph{semantic specification}
of the synthesis conjecture
for background theories $T$ from the SMT-LIB standard.
The notion of satisfying semantic restrictions for background theories
that are not standardized in the SMT-LIB standard are not covered by this document.

Before stating the formal definition for satisfying semantic specifications,
we require the following definitions.
Consider a synthesis conjecture $\exists f_1, \ldots, f_n \ldotp\Psi$
in background theory $T$, and let
$\alpha = ( \lambda X_1 \ldotp t_1, \ldots, \lambda X_n \ldotp t_n)$
be an assignment for functions-to-synthesize $( f_1, \ldots, f_n )$
whose grammars are $G_1, \ldots, G_n$.
Let $\Psi_{\alpha}$ be the formula:
\[
\bigwedge_{i=1}^{n} \forall X_i\ldotp f( X_i ) \teq t_i
\]
In other words, $\Psi_{\alpha}$ contains quantified formulas
that constrain the behavior of the functions to the synthesize
based on their definition in $\alpha$.
To reason about weight symbols in the definition below,
we say that substitution $\sigma$ is \emph{consistent with $\alpha$} if
it is of the form:
\[
\{ \wsym{f_i}{W} \mapsto k \mid \wgenerates{G_i}{W}{k}{t_i} \}
\]
In other words, $\sigma$ is consistent with $\alpha$ if it replaces
weight symbols for functions to synthesize $f_i$
with only valid interpretations based on their bodies $t_i$ in $\alpha$.

We now state the formal definition for satisfying semantic specifications
based on the above definitions.
We say that $\alpha$
satisfies the semantic restrictions for conjecture
$\exists f_1, \ldots, f_n \ldotp\Psi$
if $(\Psi_{\alpha} \wedge \Psi) \cdot \sigma$ is $T$-valid formula
for some substitution $\sigma$ that is consistent with $\alpha$.
The formal definition for $T$-valid here
corresponds to the definition given by SMT-LIB,
for details see Section 5 of~\cite{BarFT-RR-17}.
This definition covers cases where $f_1, \ldots, f_n$
have recursive definitions or references to forward declarations,
which is why their definitions are explicitly given in the definition of
$\Psi_{\alpha}$.

\subsubsection{Satisfying Semantic Specifications for Optimization Queries}
We extend the notion of semantic correctness to responses to optimization
queries of the form:
\[
\paren{\optsynthkwd\mbox{ }\paren{s_1 \ldots s_j}\mbox{ } A_1 \ldots A_k}
\]
Recall from \cref{sec:output} that a successful response to this command
contains:
\begin{enumerate}
\item A tuple of terms $(c_1, \ldots, c_j)$ corresponding to the valuation of terms $(s_1, \ldots, s_j)$, and
\item A tuple of functions $( \lambda X_1 \ldotp t_1, \ldots, \lambda X_n \ldotp t_n)$
corresponding to a solution for functions-to-synthesize $f_1, \ldots, f_n$.
\end{enumerate}
Let $\exists f_1, \ldots, f_n \ldotp\Psi$ be the synthesis conjecture
for the current state, and let $\Psi_{o}$ be the formula:
\[
\bigwedge_{i=1}^{j} s_j \teq c_j
\]
A solution of the above form for an optimization query
satisfies semantic specifications if only if 
$( \lambda X_1 \ldotp t_1, \ldots, \lambda X_n \ldotp t_n)$
satisfies the semantic specifications
for the synthesis conjecture $\exists f_1, \ldots, f_n. \Psi \wedge \Psi_{o}$.

Notice that the attributes $A_1, \ldots, A_k$ specify an ordering on the
solution, but do not impact its correctness.

\section{Calling Oracles}%
\label{sec:oracleimplementations}

In the following, we formalize the interface between the
synthesis solver and an external binary, which we call an \emph{oracle}.
An oracle can be queried either via command line
or using text files. If the oracle must be queried via text files, it is declared with the attribute $\texttt{:file}$.
If the oracle must be queried via command line, it is declared with the attribute $\texttt{:command-line}$.
The expected interface for an oracle queried via command line is given by the following definition. In absence of either attribute, 
the default assumption is that the oracle must be queried via command line.

The expected interface for an oracle queried via command line and text files are given by the following definitions:

\begin{definition}
An oracle \emph{implements an interface via command line} with input
$(\sigma_1 \times \ldots \times \sigma_n)$ and output
$(\tau_1 \times \ldots \times \tau_m)$ if
it has the following behavior.
Let ${\tt oracle}$ be the name of the binary corresponding to the oracle.
Assume the oracle is queried by command line is executed with the command
\[
{\tt oracle \,\, v_1 \ldots v_n}
\]
where $v_1, \ldots, v_n$ denote values\footnote{
A \emph{value} of type $T$ 
coincides with a term generated by the SyGuS grammar term
$\paren{\constantkwd\mbox{ }T}$.
For example, the values of type ${\tt Int}$
have syntax ${\tt N}$ or ${\tt (- N)}$ where ${\tt N}$ is a numeral.
The values for function types include all closed lambda terms.
} of sorts $\sigma_1, \ldots, \sigma_n$
written using the syntax for values described in this document.
When the above command is executed,
it returns the text on the standard output channel
\[
{\tt (w_{1} \ldots w_{m})}
\]
where $w_1, \ldots, w_n$ are values of sorts $\tau_1, \ldots, \tau_m$
written using the syntax for values described in this document.
\qed
\end{definition}

\begin{definition}
An oracle \emph{implements an interface via text files} with input
$(\sigma_1 \times \ldots \times \sigma_n)$ and output
$(\tau_1 \times \ldots \times \tau_m)$ if
it has the following behavior.
Let ${\tt oracle}$ be the name of the binary corresponding to the oracle.
Let \textit{textfile.query}
be a file containing the text
\[
{\tt (v_1 \ldots v_n)}
\]
where ${\tt v_1, \ldots, v_n}$ denote values of sorts $\sigma_1, \ldots, \sigma_n$
written using the syntax for values described in this document, similar to above.
When the oracle is executed with the command
\[
{\tt oracle \,\, textfile.query}
\]
it returns the text on the standard output channel
\[
{\tt (w_{1} \ldots w_{m})}
\]
where ${\tt w_1, \ldots, w_n}$ denote values of sorts $\tau_1, \ldots, \tau_m$
written using the syntax for values described in this document.
\qed
\end{definition}

In other words, in both cases, the synthesis solver and the external binary
communicate with one another where the text provided to the binary contains
a tuple of values corresponding to input to the binary, and the output
from the binary is a tuple of terms indicating the output of the binary,
which is printed on the standard output channel.
It is expected that the synthesis solver parses these values to
ascertain the result of the query.

It is important to note that the text file or command string is expected to be
a tuple of values only.
In some cases, 
the syntax for those values may involve symbols that are user-defined.
As an example, consider an oracle
implementing an interface whose input takes a user-defined datatype.
The definition of that datatype is \emph{not} provided as a preamble
to the given tuple of values.
This means that an external binary is responsible for
parsing the given text file, even when it uses symbols whose definitions
are not self-contained in that text file/command string. 
It is assumed that
the user has ensured the syncronization of the oracle specification
and the input format accepted by the binary.

It is the responsibility of the user to specify oracles in
SyGuS commands that implement interfaces of the appropriate sorts.
If this is not the case, the synthesis solver may choose to ignore the output
of the oracle or terminate with an error.

For example, consider an oracle that implements the interface declared with the following command:\\\\
$\paren{\constraintoraclekwd\mbox{ } 
\paren{\paren{x_1\mbox{ }\sigma_1} \ldots \paren{x_n\mbox{ }\sigma_n}}\mbox{ }
\paren{\paren{y_{1}\mbox{ }\tau_{1}} \ldots \paren{y_{m}\mbox{ }\tau_{m}}}  \mbox{ } t\mbox{ } N :file}$ \\

The text file must contain a list of values $\paren{v_1 \ldots v_n}$ for the input parameters $x_1 \ldots x_n$, and $v_i$ must have the same sort as $\sigma_i$.
The oracle will return a list of values $\paren{w_{1} \ldots w_{m}}$, corresponding to the output parameters $y_{1} \ldots y_{m}$, and $w_i$ must have the same sort as $\tau_i$.
The solver calls the oracle with the command ${\tt N \,\,textfile.query}$, where ${\tt N}$ is the name of the oracle implememtation as specified in the oracle constraint declaration.

\section{Attributes}
\label{sec:attr}

In this section, we standardize attributes that are specific to
the features introduced in this document.

\subsection{Weights}%
\label{sec:attr-weights}

\begin{enumerate}
\item The attribute $\texttt{:default}$
expects an integer attribute value $n$.
It specifies that the default weight value of declared weight keyword is $n$.
\end{enumerate}
\noindent
Recall from \cref{sec:weight-semantics} that the default value
for weight keywords is $0$.
On the other hand, the command:
\begin{lstlisting}[language=SyGuS-Desugar]
(declare-weight numOps :default 1)
\end{lstlisting}
declares a weight keyword $\texttt{numOps}$ whose default weight value is $1$.

\subsection{Oracle communication}
\label{sec:attr-oracles}
\begin{enumerate}
\item The attribute $\texttt{:file}$
specifies that an oracle implements an interface via text files, as described in \cref{sec:oracleimplementations}.
\item The attribute $\texttt{:command-line}$
specifies that an oracle implements an interface via command line, as described in \cref{sec:oracleimplementations}.
\end{enumerate}

For example, the following oracle function declaration specifies that the oracle should be queried via text files:
\begin{lstlisting}[language=SyGuS-Desugar]
(declare-oracle-fun (Int) Int binaryname :file)
\end{lstlisting}

\subsection{Objectives}%
\label{sec:attr-objectives}

In this section, we standardize attributes that pertain to specifying objectives of the $\optsynthkwd$ command.
We classify two kinds of attributes.
The first pertain to the orderings on values,
which are supplied in the first argument of $\optsynthkwd$ command.
The second pertain to orderings on the tuple itself.
Below, we use $\succ$ to denote orderings on values or tuples of values,
such that when $e_1 \succ e_2$, we have that $e_1$ is a more preferred solution.

\paragraph{Value Orderings}
\begin{enumerate}
\item The attribute $\texttt{:max}$ specifies that the ordering on values 
$\succ$ for a given term $t$ should be maximized according to the default ordering
of the type of $t$.
The specification is well-defined only if the type of $t$ has a default ordering.
We assume the default ordering $>$ is used for reals and integers, such that
$v_1 \succ v_2$ if $v_1 > v_2$.
Orderings on other types are not standardized by this document.
Note that this attribute expects no provided attribute value.
\item The attribute $\texttt{:min}$ is dual to $\texttt{:max}$, such that that ordering $\succ$
should minimize the value based on the default ordering of its type.
Like $\texttt{:max}$, this attribute expects no provided attribute value.
\end{enumerate}

\paragraph{Tuple Orderings}

\begin{enumerate}
\item The attribute $\texttt{:lexico}$ specifies that the ordering on tuples is lexicographic.
In particular, assuming orderings $\succ_1, \ldots, \succ_n$ on the components of $n$-tuples,
$(c_1, \ldots, c_n) \succ (d_1, \ldots, d_n)$ holds if there exists a $0 \leq j \leq n$
such that $c_i = d_i$ for $i<j$ and $c_j \succ_j d_j$.
This attribute expects no provided attribute value.
\end{enumerate}

\noindent
For example, consider the command:
\begin{lstlisting}[language=SyGuS-Desugar]
(optimize-synth ((! x :min) (! y :max)) :lexico)
\end{lstlisting}
This indicates that the objective is to first minimize $x$, and then maximize $y$.
Hence, in this example, solutions in response to this command are ordered such that
$(0,3) \succ (1,3) \succ (1,0)$.

\section{Examples}%
\label{sec:examples}


\begin{example}[Linear Arithmetic with Constant Coefficients]
Consider the following example:
\begin{lstlisting}[language=SyGuS]
(set-logic LIA)
(synth-fun f ((x Int) (y Int)) Int
  ((I Int) (Ic Int))
  ((I Int (0 1 x y
           (+ I I)
           (* Ic I)))
   (Ic Int (0 1 2 (- 1) (- 2)))))
(declare-var x Int)
(declare-var y Int)
(constraint (= (f x y) (* 2 (+ x y))))
(check-synth)
\end{lstlisting}
In this example, the logic is set 
to linear integer arithmetic.
The grammar of the function-to-synthesize $f$
has two non-terminals, ${\tt I}$ and ${\tt Ic}$.
What is notable in this example is that
the grammar for $f$ includes a rule for ${\tt I}$ whose right hand side 
is the term ${\tt(*\ Ic\ I)}$. 
If a term of this form were to appear
in a constraint, 
then it would not be allowed since it is the
multiplication of two non-constant terms and thus is not allowed by the input logic ${\tt LIA}$.
However, 
by the definition of our restrictions on grammars in \cref{ssec:logicr-grammars},
this grammar \emph{is allowed},
since all closed terms that the grammar generates are allowed by linear arithmetic.
A possible correct response for this example from a synthesis solver is:
\begin{lstlisting}[language=SyGuS]
(
  (define-fun f ((x Int) (y Int)) Int (+ (* 2 x) (* 2 y)))
)
\end{lstlisting}
\end{example}

\begin{example}[Datatypes with Linear Arithmetic]
Consider the following example:
\begin{lstlisting}[language=SyGuS]
(set-logic DTLIA)
(declare-datatype List
  ((nil)
   (cons (head Int) (tail List))))
(synth-fun f ((x List)) Int
  ((I Int) (L List) (B Bool))
  ((I Int (0 1 
           (head L)
           (+ I I)
           (ite B I I)))
   (L List (nil
            x
            (cons I L)
            (tail L)))
   (B Bool (((_ is nil) L)
            ((_ is cons) L)
            (= I I)
            (>= I I)))))
(constraint (= (f (cons 4 nil)) 5))
(constraint (= (f (cons 0 nil)) 1))
(constraint (= (f nil) 0))
(check-synth)
\end{lstlisting}
In this example, the logic is set to 
datatypes with linear integer arithmetic ${\tt DTLIA}$.
A datatype ${\tt List}$ is then declared,
which encodes lists of integers with two constructors ${\tt nil}$ and ${\tt cons}$.
The input contains a single-function to synthesize $f$
that takes as input a list and returns an integer.
Its grammar contains non-terminal symbols for integers, lists and Booleans,
and includes applications of constructors ${\tt nil}$ and ${\tt cons}$,
selectors ${\tt head}$ and ${\tt tail}$, and
discriminators ${\tt (\_\ is\ nil)}$ and ${\tt (\_\ is\ cons)}$.
A possible correct response for this example from a synthesis solver is:
\begin{lstlisting}[language=SyGuS]
(
  (define-fun f ((x List)) Int (ite ((_ is nil) x) 0 (+ 1 (head x))))
)
\end{lstlisting}
In other words, a possible solution for $f$ returns zero whenever its
argument is the empty list ${\tt nil}$, 
and returns one plus the head of its argument otherwise.
\end{example}

\begin{example}[Bit-Vectors with Concatenation and Extraction]
Consider the following example:
\begin{lstlisting}[language=SyGuS]
(set-logic BV)
(synth-fun f ((x (_ BitVec 32))) (_ BitVec 32)
  ((BV32 (_ BitVec 32)) (BV16 (_ BitVec 16)))
  ((BV32 (_ BitVec 32) (#x00000000 #x00000001 #xFFFFFFFF
                        x
                        (bvand BV32 BV32)
                        (bvor BV32 BV32)
                        (bvnot BV32)
                        (concat BV16 BV16)
                        ))
   (BV16 (_ BitVec 16) (#x0000 #x0001 #xFFFF
                        (bvand BV16 BV16)
                        (bvor BV16 BV16)
                        (bvnot BV16)
                        ((_ extract 31 16) BV32)
                        ((_ extract 15 0) BV32)))))
(constraint (= (f #x0782ECAD) #xECAD0000))
(constraint (= (f #xFFFF008E) #x008E0000))
(constraint (= (f #x00000000) #x00000000))
(check-synth)
\end{lstlisting}
The this example, the logic is set to bit-vectors ${\tt BV}$.
A single function-to-synthesize $f$ is given that takes a
bit-vector of bit-width 32 as input and returns a bit-vector of the same width as output.
Its grammar involves non-terminals whose sorts are bit-vectors of
bit-width 32 and 16.
The semantics of the operators in this example are defined in the SMT-LIB standard.
In particular,
the operator ${\tt concat}$ concatenates its two arguments,
and the indexed operator ${\tt (\_\ extract\ n\ m)}$
returns a bit-vector containing bits $n$ through $m$ of its
argument, where $n \geq m$.
A possible correct response for this example from a synthesis solver is:
\begin{lstlisting}[language=SyGuS]
(
  (define-fun f ((x (_ BitVec 32))) (_ BitVec 32)
     (concat ((_ extract 15 0) x) #x0000))
)
\end{lstlisting}
In other words, a possible solution for $f$ 
returns the concatenation of bits $15$ to $0$ of its argument
with the bit-vector ${\tt \#x0000}$.
\end{example}

\begin{example}[Grammars with Defined Symbols, Forward Declarations and Recursion]
Consider the following example:
\begin{lstlisting}[language=SyGuS]
(set-logic LIA)
(set-feature :fwd-decls true)
(set-feature :recursion true)
(define-fun x_plus_one ((x Int)) Int (+ x 1))
(synth-fun f ((x Int)) Int
  ((I Int))
  ((I Int (0 1 x (x_plus_one I)))))
(define-fun fx_plus_one ((x Int)) Int (+ (f x) 1))
(synth-fun g ((x Int)) Int
  ((I Int))
  ((I Int (0 1 x (fx_plus_one I)))))
(synth-fun h ((x Int)) Int
  ((I Int) (B Bool))
  ((I Int (0 1 x (- I 1) (+ I I) (h I) (ite B I I)))
   (B Bool ((= I I) (> I I)))))
(declare-var y Int)
(constraint (= (h y) (- (g y) (f y))))
(check-synth)
\end{lstlisting}
This example contains three well-formed $\synthfunkwd$ commands.
The first one for function $f$ contains an application of a macro ${\tt x\_plus\_one}$,
which abbreviates adding one to its argument.
This grammar is allowed in the default SyGuS logic
and in this example.
The grammar for $g$ contains an application of a macro ${\tt fx\_plus\_one}$,
whose expanded form contains a previously declared function-to-synthesize $f$.
This grammar is allowed since the feature $\ffwddeclskwd$ is enabled in this example.
The grammar for $h$ contains a rule that contains an application of $h$ itself.
This grammar is allowed since $\frecursionkwd$ is enabled in this example.
A possible correct response for this example from a synthesis solver is:
\begin{lstlisting}[language=SyGuS]
(
  (define-fun f ((x Int)) Int x)
  (define-fun g ((x Int)) Int (fx_plus_one x))
  (define-fun h ((x Int)) Int 1)
)
\end{lstlisting}
Notice the above definitions for $f$, $g$, and $h$ are given 
in the order in which they were introduced via $\synthfunkwd$ commands
in the input.
\end{example}

\begin{example}[Programming by Examples (PBE) with Strings]
Consider the following example:
\begin{lstlisting}[language=SyGuS]
(set-logic PBE_SLIA)
(synth-fun f ((fname String) (lname String)) String
  ((y_str String) (y_int Int))
  ((y_str String (" " fname lname
                  (str.++ y_str y_str)
                  (str.replace y_str y_str y_str)
                  (str.at y_str y_int)
                  (str.from_int y_int)
                  (str.substr y_str y_int y_int)))
     (y_int Int (0 1 2
                 (+ y_int y_int)
                 (- y_int y_int)
                 (str.len y_str)
                 (str.to_int y_str)
                 (str.indexof y_str y_str y_int)))))
(constraint (= (f "Nancy" "FreeHafer") "Nancy FreeHafer"))
(constraint (= (f "Andrew" "Cencici") "Andrew Cencici"))
(constraint (= (f "Jan" "Kotas") "Jan Kotas"))
(constraint (= (f "Mariya" "Sergienko") "Mariya Sergienko"))
(check-synth)
\end{lstlisting}
In this example, the logic is set to the SyGuS-specific logic
${\tt PBE\_SLIA}$, indicating strings with linear integer arithmetic
where constraints are limited to a conjunction of PBE equalities.
In this example, four constraints are given, each of which meet the
restriction of being a PBE equality.
A possible correct response for this example from a synthesis solver is:
\begin{lstlisting}[language=SyGuS]
(
  (define-fun f ((fname String) (lname String)) String
    (str.++ fname (str.++ " " lname)))
)
\end{lstlisting}
\end{example}

\begin{example}[Weights]
Consider the following example:
\begin{lstlisting}[language=SyGuS]
(set-logic NIA)
(set-feature :weights true)
(declare-weight numX)
(synth-fun f ((x Int)) Int
  ((I Int))
  ((I Int (0 1
           (! x :numX 1)
           (+ I I)
           (! (* x x) :numX 2)))))
(constraint (= (_ numX f) 3))
(check-synth)
\end{lstlisting}
In this example, a weight keyword $\texttt{numX}$ has been declared.
A function-to-synthesize $f$ has been declared, whose production
rules include annotations.
These annotations count the number occurrences of $x$, where recall
that the default value for unannotated terms with respect to the given weight is $0$.
A constraint is given that states that the weight of $f$ with respect to
the weight keyword $\texttt{numX}$ must be $3$.
A possible correct response for this example from a synthesis solver is:
\begin{lstlisting}[language=SyGuS]
(
  (define-fun f ((x Int)) Int (+ x (* x x)))
)
\end{lstlisting}
\end{example}

\begin{example}[Weight with Multiple Interpretations]
Consider the following example:
\begin{lstlisting}[language=SyGuS]
(set-logic LIA)
(set-feature :weights true)
(declare-weight numI)
(synth-fun f ((x Int)) Int
  ((I Int))
  ((I Int (0 1 x
           (+ x 1)
           (! (- I) :numI 1)
           (! (+ I I) :numI 2)))))
(define-fun numRulesForF () Int (+ (_ numI f) 1))
(declare-var x Int)
(constraint (and (> (f x) x) (< numRulesForF 3)))
(check-synth)
\end{lstlisting}
In this example, a weight keyword $\texttt{numI}$ has been declared,
which counts the number of occurrences of the non-terminal \texttt{I} that
were used in applications of the production rules that generated the body of $f$.
The function $\texttt{numRulesForF}$ is defined to be the weight of $f$ with
respect to this keyword plus one, or in other words, the number of
rules used for generating the body of $f$.
A possible correct response for this example from a synthesis solver is:
\begin{lstlisting}[language=SyGuS]
(
  (define-fun f ((x Int)) Int (+ x 1))
)
\end{lstlisting}
In this example,
the body \texttt{(+ x 1)} given for $f$ can be generated in multiple ways from the given grammar.
In particular, it could be generated directly from the start symbol,
or e.g. via the sequence \texttt{I}, \texttt{(+ I I)}, \texttt{(+ x I)}, \texttt{(+ x 1)}.
Thus, when the body of $f$ is interpreted as \texttt{(+ x 1)},
the interpretation of \texttt{(+ (\_ numI f) 1)} may either be $1$ or $3$.
The constraint \texttt{(< numRulesForF 3)} is satisfied in the former interpretation,
and hence the solution for $f$ satisfies all input constraints.
\end{example}

\begin{example}[Optimization with Weights]
Consider the following example:
\begin{lstlisting}[language=SyGuS]
(set-logic LIA)
(set-feature :weights true)
(declare-weight branches)
(synth-fun f ((x Int)) Int
   ((I Int) (B Bool))
   ((I Int (0 1 x (+ I I) (! (ite B I I) :branches 1)))
    (B Bool ((>= I I) (= I I)))))
(declare-var x Int)
(constraint (= (f x) (ite (= x 0) 0 x)))
(optimize-synth ((! (_ branches f) :min)))
\end{lstlisting}
In this example, a weight keyword $\texttt{branches}$ has been declared.
The grammar of function $\texttt{f}$ has been annotated such that
$\texttt{branches}$ corresponds to the number of $\texttt{ite}$ terms in its body.
An optimization query is made, where
the objective is to minimize the number of branches in $\texttt{f}$.
This is specified by a term annotation on $\texttt{(\_ branches f)}$.
A possible correct response for this example is:
\begin{lstlisting}[language=SyGuS]
(
(1)
(define-fun f ((x Int)) Int (ite (= x 0) 0 x))
)
\end{lstlisting}
Another possible response for this example is:
\begin{lstlisting}[language=SyGuS]
(
(0)
(define-fun f ((x Int)) Int x)
)
\end{lstlisting}
In both cases, the value of term $\texttt{(\_ branches f)}$ was given
in the solution that was consistent with the number of
$\texttt{ite}$ terms in the body of $\texttt{f}$.
Hence, both responses are correct.
The latter solution gives a smaller value for this term,
and hence it is preferred based on the specified objective.

Notice that in this example,
no attribute is provided for the $\texttt{optimize-synth}$ command,
meaning that no ordering is specified on the tuple passes as argument
to this command.
For details on the default semantics for objectives, see \cref{sec:attr-objectives}.
In the next example, we show how ordering on tuples can be specified.
\end{example}

\begin{example}[Lexicographic Optimization]
Consider the following example:
\begin{lstlisting}[language=SyGuS]
(set-logic LIA)
(synth-fun f ((x Int)) Int
   ((I Int) (B Bool))
   ((I Int (0 1 x (+ I I) (ite B I I)))))
(declare-var x Int)
(constraint (or (= (f x) 1) (= (f x) x)))
(optimize-synth ((! (f 0) :max) (! (f 100) :max)) :lexico)
\end{lstlisting}
This example also contains an optimization query.
The objective of the query is to optimize the values
of terms $\texttt{(f 0)}$ and $\texttt{(f 100)}$.
Moreover, the ordering on tuples of these terms is lexicographic
with respect to the orderings specified for each term position.
In particular, this means that the best solution for this
example is one that first maximizes $\texttt{(f 0)}$ and then
maximizes $\texttt{(f 100)}$.
A possible correct response for this example is:
\begin{lstlisting}[language=SyGuS]
(
(1 1)
(define-fun f ((x Int)) Int 1)
)
\end{lstlisting}
Another possible response for this example is:
\begin{lstlisting}[language=SyGuS]
(
(0 100)
(define-fun f ((x Int)) Int x)
)
\end{lstlisting}
The first example is preferred, since the first value in the first
solution is greater than the first value in the second,
since the command specified that solutions are ordered lexicographically.
\end{example}

\begin{example}[Single Invariant Synthesis (Inv) over Linear Integer Arithmetic]
Consider the following imperative program in a C-like language:
\begin{lstlisting}[language=C-Like]
int x, y;
assume (5 <= x && x <= 9);
assume (1 <= y && y <= 3);
while (*) {
  x += 2;
  y += 1;
}
assert (y < x);
\end{lstlisting}
The problem of synthesizing an invariant to verify this program,
can be expressed in SyGuS as follows:
\begin{lstlisting}[language=SyGuS]
(set-logic Inv_LIA)
(synth-fun inv-f ((x Int) (y Int)) Bool)
(define-fun pre-f ((x Int) (y Int)) Bool 
  (and (<= 5 x) (<= x 9) (<= 1 y) (<= y 3)))
(define-fun trans-f ((x Int) (y Int) (xp Int) (yp Int)) Bool 
  (and (= xp (+ x 2)) (= yp (+ y 1))))
(define-fun post-f ((x Int) (y Int)) Bool (< y x))
(inv-constraint inv-f pre-f trans-f post-f)
(check-synth)
\end{lstlisting}
In this example, the logic is set to the SyGuS-specific logic
\code|Inv_LIA|, indicating linear integer arithmetic
where constraints are limited to the invariant synthesis problem for
a single invariant-to-synthesize.
Since this example contains only linear integer arithmetic terms,
a single predicate-to-synthesize \code|inv-f|,
and only introduces constraints via a single $\constraintinvkwd$ command,
it meets the restrictions of the logic.
A possible correct response for this example from a synthesis solver is:
\begin{lstlisting}[language=SyGuS]
(
  (define-fun inv-f ((x Int) (y Int)) Bool (> x y))
)
\end{lstlisting}
\end{example}

\begin{example}[CHCs over Linear Integer Arithmetic with a Single Predicate to Synthesize]
The verification problem from the previous example can also be expressed as a system of CHCs,
which can be encoded in SyGuS as follows:
\begin{lstlisting}[language=SyGuS]
(set-logic CHC_LIA)
(synth-fun inv-f ((x Int) (y Int)) Bool)
(chc-constraint ((x Int) (y Int))
  (and (<= 5 x) (<= x 9) (<= 1 y) (<= y 3))
  (inv-f x y))
(chc-constraint ((x Int) (y Int) (xp Int) (yp Int))
  (and (inv-f x y) (= xp (+ x 2)) (= yp (+ y 1)))
  (inv-f xp yp))
(chc-constraint ((x Int) (y Int))
  (and (inv-f x y) (not (< y x)))
  false)
(check-synth)
\end{lstlisting}
In this example, the logic is set to the SyGuS-specific logic
\code|CHC_LIA|, indicating linear integer arithmetic
where constraints are limited to constrained Horn clauses.
Since this example contains only linear integer arithmetic terms,
only predicate symbols to be synthesized (i.e., \code|inv-f|,
only introduces valid CHCs via $\constraintchckwd$ command, and
only a single CHC query,
it meets the restrictions of the logic.
As for the previous example, a possible correct response for this example from a synthesis solver is:
\begin{lstlisting}[language=SyGuS]
(
  (define-fun inv-f ((x Int) (y Int)) Bool (> x y))
)
\end{lstlisting}
\end{example}

\begin{example}[CHCs over Linear Integer Arithmetic with Multiple Predicates to Synthesize]
Consider the following imperative program in a C-like language:
\begin{lstlisting}[language=C-Like]
int x;
int y = x, n = 0;
while (*) {
  x += 1;
  n += 1;
}
x *= 2;
while (n != 0) {
  x -= 2;
  n -= 1;
}
assert (x == 2 * y);
\end{lstlisting}
The verification problem for this program is more naturally expressed as a system of CHCs
using the \code|CHC| logic in SyGuS instead of the \code|Inv| logic:
\begin{lstlisting}[language=SyGuS]
(set-logic CHC_LIA)
(synth-fun inv1 ((x Int) (y Int) (n Int)) Bool)
(synth-fun inv2 ((x Int) (y Int) (n Int)) Bool)
(chc-constraint ((x Int) (y Int) (n Int))
  (and (= y x) (= n 0))
  (inv1 x y n))
(chc-constraint ((x Int) (y Int) (n Int) (xp Int) (np Int))
  (and (inv1 x y n) (= xp (+ x 1)) (= np (+ n 1)))
  (inv1 xp y np))
(chc-constraint ((x Int) (y Int) (n Int) (xp Int))
  (and (inv1 x y n) (= xp (* 2 x)))
  (inv2 xp y n))
(chc-constraint ((x Int) (y Int) (n Int) (xp Int) (np Int))
  (and (inv2 x y n) (not (= n 0)) (= xp (- x 2)) (= np (- n 1)))
  (inv2 xp y np))
(chc-constraint ((x Int) (y Int) (n Int))
  (and (= n 0) (inv2 x y n) (not (= x (* 2 y))))
  false)
(check-synth)
\end{lstlisting}
In this example, the logic is set to the SyGuS-specific logic
\code|CHC_LIA|, indicating linear integer arithmetic
where constraints are limited to constrained Horn clauses.
Since this example contains only linear integer arithmetic terms,
only predicate symbols to be synthesized (i.e., \texttt{inv1} and \texttt{inv2}),
only introduces valid CHCs via $\constraintchckwd$ command, and
only a single CHC query,
it meets the restrictions of the logic.
A possible correct response for this example from a synthesis solver is:
\begin{lstlisting}[language=SyGuS]
(
  (define-fun inv1 ((x Int) (y Int) (n Int)) Bool (= x (+ y n)))
  (define-fun inv2 ((x Int) (y Int) (n Int)) Bool (= x (* 2 (+ y n))))
)
\end{lstlisting}
\end{example}

\begin{example}[PBE with Oracle Function Symbols]
Oracle function symbols are used to represent black-box functions within specifications. Consider a PBE problem where the user has an external oracle and wishes to synthesize a summary of the external oracle for a specific set of inputs. This can be encoded
using an oracle function symbol (note this is syntactic sugar for an oracle assumption and an uninterpreted function):
\begin{lstlisting}[language=SyGuS]
(set-logic BV)
(set-feature :oracles true) 
(synth-fun f ((x (_ BitVec 64))) (_ BitVec 64))
(declare-oracle-fun target ((_ BitVec 64)) (_ BitVec 64) binaryname)

(constraint (= (f #x28085a970e13e12c) (target #x28085a970e13e12d)))
(constraint (= (f #xbe5341bebd2a0749) (target #xbe5341bebd2a0749)))
(constraint (= (f #xe239460eed2cc34e) (target #xe239460eed2cc34f)))
(constraint (= (f #xac5b1b5e9b236b10) (target #xac5b1b5e9b236b11)))
(constraint (= (f #x4069a4c7173e1786) (target #x4069a4c7173e1786)))
(constraint (= (f #x39419062091119a6) (target #x39419062091119a6)))
(constraint (= (f #x49aeeca628644ee0) (target #x49aeeca628644ee0)))
(constraint (= (f #x75e5bc2a07c77c97) (target #x75e5bc2a07c77c97)))
(constraint (= (f #x4c5ee4be98c5ee7d) (target #x4c5ee4be98c5ee7d)))
(constraint (= (f #xcd67bd5beaac575e) (target #xcd67bd5beaac575e)))
(check-synth)
\end{lstlisting}
In this way, a PBE problem can be encoded without the user knowing the correct behavior of the target function explicitly. 
Instead, the target function is implemented by the oracle \lstinline{binaryname}, 
and the solver can call \lstinline{binaryname} with an input given in a constraint 
and \lstinline{binaryname} returns the result of applying \lstinline{target} to the input.
\end{example}

\begin{example}[Synthesis with Oracle Assumptions]
In addition to their use in oracle function symbols, oracle assumptions can be used to relax a synthesis specification in other ways. Consider the following synthesis problem. The oracle, when called, generates an assumption that relaxes the values of x that the specification must hold for, i.e., when $x < z$ the specification need not hold. This could be used, for example, in invariant synthesis if an oracle can determine that certain parts of the state-space are unreachable from the initial states and backward unreachable from the violation of the post-condition. This could be particularly useful if the specified transition relation is over-approximate.

\begin{lstlisting}[language=SyGuS]
(set-logic LIA)
(set-feature :oracles true) 
(synth-fun inv-f ((x Int)) Bool)
(declare-var x Int)
(declare-var x! Int)

(define-fun pre-f ((x Int)) Bool (= (mod x 2) 0))
(define-fun trans-f ((x Int)(x! Int)) Bool (or (= x! x) (= x! (* x (x-1)))))
(define-fun post-f ((x Int)) Bool (= (mod x 4) 0)

(constraint (=> (pre-f x) (inv-f x)))
(constraint (=> (and (inv-f x) (trans-f x x!)) (inv-f x!)))
(constraint (=> (inv-f x) (post-f x)))

(oracle-assume ((inv-candidate (-> Int Bool))) ((z Int)) binaryname
  (>= x z))

(check-synth)
\end{lstlisting}
\end{example}

\begin{example}[Synthesis with Oracle Constraints and Oracle Function Symbols]

As well as PBE style problems, oracle function symbols can be used to incorporate other verification engines via correctness oracles. Here, an oracle function symbol must return ``true'' before the synthesised function is valid. However, a black-box correctness oracle along does not provide much information to the synthesiser. This oracle can be supplemented with an oracle constraint. Here the oracle constraint provides a fresh positive witness whenever it is called, constraining the space of possible candidate functions. 

\begin{lstlisting}[language=SyGuS]
(set-logic LIA)
(set-feature :oracles true) 
(synth-fun f ((x Int)) Bool)
(declare-oracle-fun isCorrect ((-> Int Bool)) Bool binaryname)

(constraint (isCorrect f))

(oracle-constraint () ((x Int)) ((z Bool)) binaryname2
  (=> (f x) z))

(check-synth)
\end{lstlisting}
\end{example}

\bibliographystyle{plain}
\bibliography{references}

\newpage
\begin{appendix}

\section{Reserved Words}%
\label{apx:reserved}

A \emph{reserved word} is any of the 
literals from \cref{ssec:literals},
or any of the following keywords:\\
\code|!|,
\code|_|,
$\checksynthkwd$,
$\constantkwd$,
$\constraintchckwd$,
$\constraintkwd$,
$\corrcexoracledeckwd$,
$\corroracledeckwd$,
$\dtdeclkwd$,
$\dtsdeclkwd$,
$\oracledeckwd$,
$\sortdeclkwd$,
$\vardeclkwd$,
$\weightdeclkwd$,
$\fundefkwd$,
$\sortdefkwd$,
$\existskwd$,
$\forallkwd$,
$\constraintinvkwd$,
$\letkwd$,
$\optsynthkwd$,
$\assumeoraclekwd$,
$\constraintoraclekwd$,
$\cexoracledeckwd$,
$\iooracledeckwd$,
$\memoracledeckwd$,
$\negwitnessoracledeckwd$,
$\poswitnessoracledeckwd$,
$\setfeaturekwd$,
$\setinfokwd$,
$\setlogickwd$,
$\setoptkwd$,
$\synthfunkwd$, and
$\varkwd$.

\section{Reference Grammars}%
\label{apx:ref-grammars}

In this section, for convenience, we provide the concrete syntax for
grammars that generate exactly the set of terms 
that belong to SMT-LIB logics of interest for a fixed set of free variables.
In particular, this means that each of the following $\synthfunkwd$
commands are equivalent to those in which no grammar is provided.
Note this is not intended to be a complete list of logics.
In particular, we focus on logics that include a single background theory
whose sorts are not parametric.
Each of these grammars are derived
based on the definition of logics and theories
described in the theory and logic declaration documents
available at \url{www.smt-lib.org}.

For each grammar, we omit 
the predicate symbols that are shared by all logics
according to the SMT-LIB standard, that is,
those included in the core theory described in \cref{ssec:smt-logic},
which includes Boolean connectives and equality.
We provide the grammar
for a single function over one of the sorts in the logic,
and assume it has one variable in its argument list
for each non-Boolean sort in the grammar.

\subsection{Integer Arithmetic}

The following grammar for $f$
generates exactly the integer-typed terms in the logic of
linear integer arithmetic (LIA) with one free integer variable ${\tt x}$.
\begin{lstlisting}[language=SyGuS]
(set-logic LIA)
(synth-fun f ((x Int)) Int
  ((y_int Int) (y_const_int Int) (y_bool Bool))
  ((y_int Int (y_const_int
               (Variable Int)
               (- y_int)
               (+ y_int y_int)
               (- y_int y_int)
               (* y_const_int y_int)
               (* y_int y_const_int)
               (div y_int y_const_int)
               (mod y_int y_const_int)
               (abs y_int)
               (ite y_bool y_int y_int)))
   (y_const_int Int ((Constant Int)))
   (y_bool Bool ((= y_int y_int)
                 (> y_int y_int)
                 (>= y_int y_int)
                 (< y_int y_int)
                 (<= y_int y_int)))))
\end{lstlisting}
Above, \code|div| denotes integer division,
\code|mod| denotes integer modulus and \code|abs| denotes
the absolute value function.
Positive integer constants and zero are written using the syntax
for numerals $\intconst$. Negative integer constants
are written as the unary negation of a positive integer constant.

The following grammar for $g$ generates exactly the integer-typed terms in the logic of
non-linear integer arithmetic (NIA)
with one free integer variable $x$.
\begin{lstlisting}[language=SyGuS]
(set-logic NIA)
(synth-fun g ((x Int)) Int
  ((y_int Int) (y_bool Bool))
  ((y_int Int ((Constant Int)
               (Variable Int)
               (- y_int)
               (+ y_int y_int)
               (- y_int y_int)
               (* y_int y_int)
               (div y_int y_int)
               (mod y_int y_int)
               (abs y_int)
               (ite y_bool y_int y_int)))
   (y_bool Bool ((= y_int y_int)
                 (> y_int y_int)
                 (>= y_int y_int)
                 (< y_int y_int)
                 (<= y_int y_int)))))
\end{lstlisting}

\subsection{Real Arithmetic}

The following grammar for $f$
generates exactly the real-typed terms in the logic of
linear real arithmetic (LRA)
with one free real variable $x$.
\begin{lstlisting}[language=SyGuS]
(set-logic LRA)
(synth-fun f ((x Real)) Real
  ((y_real Real) (y_const_real Real) (y_bool Bool))
  ((y_real Real (y_const_real 
                 (Variable Real) 
                 (- y_real)
                 (+ y_real y_real)
                 (- y_real y_real)
                 (* y_const_real y_real)
                 (* y_real y_const_real)
                 (/ y_real y_const_real)
                 (ite y_bool y_real y_real)))
   (y_const_real Real ((Constant Real)))
   (y_bool Bool ((= y_real y_real)
                 (> y_real y_real)
                 (>= y_real y_real)
                 (< y_real y_real)
                 (<= y_real y_real)))))
\end{lstlisting}
Notice that positive real constants and zero can either be written 
as decimal values using the syntax $\realconst$
or as rationals, e.g. the division of two numerals
\code|(/ m n)| where $n$ is not zero.
Negative reals are written either as the unary negation of decimal value 
or a rational of the form \code|(/ (- m) n)| for numerals $m$ and $n$
where $n$ is not zero.

The following grammar for $g$
generates exactly the real-typed terms in the logic of
non-linear real arithmetic (NRA)
with one free real variable $x$.
\begin{lstlisting}[language=SyGuS]
(set-logic NRA)
(synth-fun g ((x Real)) Real
  ((y_real Real) (y_bool Bool))
  ((y_real Real ((Constant Real) 
                 (Variable Real)
                 (- y_real)
                 (+ y_real y_real)
                 (- y_real y_real)
                 (* y_real y_real)
                 (/ y_real y_real)
                 (ite y_bool y_real y_real)))
   (y_bool Bool ((= y_real y_real)
                 (> y_real y_real)
                 (>= y_real y_real)
                 (< y_real y_real)
                 (<= y_real y_real)))))
\end{lstlisting}

\subsection{Fixed-Width Bit-Vectors}

The signature of bit-vectors includes an indexed sort \code|BitVec|,
which is indexed by an integer constant that denotes its bit-width.
We show a grammar below for one particular choice of this bit-width, $32$.
We omit indexed bit-vector operators
such as extraction function \code|(_ extract m n)|,
the concatenation function \code|concat|,
since these operators are polymorphic.
For brevity,
we also omit the \emph{extended} operators of this theory
as denoted by SMT-LIB,
which includes functions like
bit-vector subtraction \code|bvsub|,
exclusive or \code|bvxor|,
signed division \code|bvsdiv|,
and predicates like
unsigned-greater-than-or-equal \code|bvuge|
and signed-less-than \code|bvslt|.
These extended operators can be seen as syntactic sugar for the
operators in the grammar below.
Example inputs that involve 
some of the omitted operators are given in \cref{sec:examples}.

\begin{lstlisting}[language=SyGuS]
(set-logic BV)
(synth-fun f ((x (_ BitVec 32))) (_ BitVec 32)
  ((y_bv (_ BitVec 32)) (y_bool Bool))
  ((y_bv (_ BitVec 32) ((Constant (_ BitVec 32))
                        (Variable (_ BitVec 32))
                        (bvnot y_bv)
                        (bvand y_bv y_bv)
                        (bvor y_bv y_bv)
                        (bvneg y_bv)
                        (bvadd y_bv y_bv)
                        (bvmul y_bv y_bv)
                        (bvudiv y_bv y_bv)
                        (bvurem y_bv y_bv)
                        (bvshl y_bv y_bv)
                        (bvlshr y_bv y_bv)
                        (ite y_bool y_bv y_bv)))
   (y_bool Bool ((bvult y_bv y_bv)))))
\end{lstlisting}
Bit-vector constants may be specified
using either the hexadecimal format $\hexconst$
or the binary format $\binaryconst$.

\subsection{Strings}

The syntax below reflects the official version 2.6 release of the theory
of Unicode strings and regular expressions~\cite{BarFT-RR-17}.
Some operators (e.g. indexed regular expression operators)
have been omitted from the grammar below for the sake of brevity.
\begin{lstlisting}[language=SyGuS]
(set-logic S)
(synth-fun f ((xs String) (xr RegLan) (xi Int)) String
  ((y_str String) (y_rl RegLan) (y_int Int) (y_bool Bool))
  ((y_str String ((Constant String)
                  (Variable String)
                  (str.++ y_str y_str)
                  (str.at y_str y_str)
                  (str.substr y_str y_int y_int)
                  (str.indexof y_str y_str y_int)
                  (str.replace y_str y_str y_str)
                  (str.from_int y_int)
                  (str.from_code y_int)
                  (ite y_bool y_str y_str)))
    (y_rl RegLan ((Constant RegLan)
                  (Variable RegLan)
                  re.none
                  re.all
                  re.allchar
                  (str.to_re y_str)
                  (re.++ y_rl y_rl)
                  (re.union y_rl y_rl)
                  (re.inter y_rl y_rl)
                  (re.* y_rl)
                  (re.+ y_rl)
                  (re.opt y_rl)
                  (re.range y_str y_str)
    (y_int Int ((Constant Int)
                (Variable Int)
                (str.len y_str)
                (str.to_int y_str)
                (str.to_code y_str)
                (ite y_bool y_int y_int)))
    (y_bool Bool ((Constant Bool)
                  (Variable Bool)
                  (str.in_re y_str y_rl)
                  (str.contains y_str y_str)
                  (str.prefixof y_str y_str)
                  (str.suffixof y_str y_str)
                  (str.< y_str y_str)
                  (str.<= y_str y_str)
                  (str.is_digit y_str)))))
\end{lstlisting}
String constants may be specified by text delimited by double quotes
based on the syntax $\stringconst$ described in \cref{ssec:literals}.
The sort \code|RegLan| denotes the regular expression sort.
Its values are all ground (i.e. variable free) regular expressions.

Notice that since the logic specified above is \code|S|.
The signature of this logic includes 
some functions involving the integer sort like \code|str.len|.
However, the above logic does not permit inputs containing
integer constants or the standard symbols of arithmetic like 
\code|+|, \code|-|, \code|>=| and so on, since the logic \code|S|
does not include the theory of integer arithmetic.
Thus in pratice, 
the string theory is frequently combined with the theory of integers
in the logic \code|SLIA|, i.e. strings with linear integer arithmetic.

\section{Language Features of SMT-LIB not Covered}%
\label{apx:not-covered}

For the purpose of self-containment,
many of the essential language features of SMT-LIB version 2.6
are redefined in this document.
However, other less essential ones are omitted.
We briefly mention other language features not mentioned in
this document.
We do not require solvers to support these features.
However, we recommend that if solvers support any of the features below,
they use SMT-LIB compliant syntax, as described briefly below.

\paragraph{Parametric Datatype Definitions}
We do not cover the concrete syntax or semantics of parametric datatypes (that
is, datatypes whose arity is non-zero) in this document.
An example of a datatype definition for a parametric list is given below.
\begin{lstlisting}
(declare-datatypes ((List 1)) (
  (par (T) ((nil) (cons (head T) (tail (List T)))))))
\end{lstlisting}
Above, the datatype ${\tt List}$ 
is given in the predeclaration with the numeral ${\tt 1}$, indicating
its arity is one.
The datatype definition that follows includes quantification on a type
parameter ${\tt T}$, where this quantification is specified using the 
${\tt par}$ keyword. Within the body of this quantification,
a usual constructor listing is given, where the type parameter ${\tt T}$
may occur free. The constructors of this datatype are ${\tt nil}$
and ${\tt cons}$.

\paragraph{Qualified Identifiers}
In SMT-LIB version 2.6,
identifiers that comprise terms
may be \emph{qualified} with a type-cast, using the keyword $\askwd$.
Type casts are required for symbols whose type is ambiguous,
such as parametric datatype constructors, e.g. the nil constructor
for a parametric list
For the parametric datatype above, 
${\tt (as\ nil\ (List\ Int))}$ and ${\tt (as\ nil\ (List\ Real))}$ 
denote the type constructors
for the empty list of integers and reals respectively.

\paragraph{Match Terms}
The SMT-LIB version 2.6 includes a ${\tt match}$ term
that case splits on the constructors of datatype terms.
For example,
the match term:
\begin{lstlisting}
(match x ((nil (- 1)) ((cons h t) h)))
\end{lstlisting}
returns negative one if the list $x$ is empty,
or the first element of $x$ (its head) if it is non-empty.

\paragraph{Recursive Functions}
The SMT-LIB version 2.6 includes a command $\recfundefkwd$
for defining symbols that involve recursion.\footnote{
Recall that recursive definitions are prohibited from macro definitions in the command
$\fundefkwd$.
}
An example, which computes the length of a (non-parametric) list 
is given below.
\begin{lstlisting}
(define-fun-rec len ((x List)) Int 
  (match x ((nil 0) ((cons h t) (+ 1 (len t))))))
\end{lstlisting}
Functions may be defined to be mutually recursive by declaring
them in a single block using the $\recfunsdefkwd$ command.
An example, which defines two predicates ${\tt isEven}$ and ${\tt isOdd}$
for determining whether a positive integer is even and odd respectively,
is given in the following.
\begin{lstlisting}
(define-funs-rec (
  (isEven ((x Int)) Bool)
  (isOdd ((x Int)) Bool)
)(
  (ite (= x 0) true (isOdd (- x 1)))
  (ite (= x 0) false (isEven (- x 1)))
))
\end{lstlisting}
Notice that recursive function definitions are not required to be terminating
(the above functions do not terminate for negative integers).
They are not even required to correspond to consistent definitions, for instance:
\begin{lstlisting}
(define-fun-rec inconsistent ((x List)) Int 
  (+ (inconsistent x) 1))
\end{lstlisting}
The semantics of recursive functions is given in 
the SMT-LIB version 2.6 standard~\cite{BarFT-RR-17}.
Each recursive function can be seen as a universally quantified constraint
that asserts that each set of values in the domain of the function
is equal to its body.


\paragraph{Additional Logics and Theories}
Several standard logics and theories are omitted from discussion in this document.
This includes the (mixed) theory of integers and reals,
the theory of floating points,
the integer and real difference logic,
and their combinations.
More details on the catalog of logics and theories in the SMT-LIB standard
is available at \url{www.smt-lib.org}.

\end{appendix}

\end{document}

%% file: listing-style.tex
\lstdefinelanguage{C-Like}[]{C}{
    keywordstyle=[3]\color{teal!90!black},
    keywords=[3]{assume, assert}
}

\lstdefinelanguage{SyGuS}{
    alsoletter={\_,-,+,*,=,>,<,:,!},
    mathescape=true,
    emptylines=0,
    texcl=true,
    morecomment=[l]{;},
    morestring=[d][]{\"},
    keywordstyle=[10]\color{blue!75!black},
    keywords=[10]{check-synth, %
                  optimize-synth, %
                  set-feature, set-logic, set-option},
    keywordstyle=[20]\color{purple},
    keywords=[20]{BV, CHC, CHC_, CHC_QF_, CHC_LIA, DTLIA, DTNIA, %
                  Inv, Inv_, Inv_QF_, Inv\_LIA, LIA, LRA, NIA, %
                  PBE, PBE_, PBE\_SLIA, QF_, S, SLIA},
    keywordstyle=[21]\color{purple!80!black},
    keywords=[21]{BitVec, Bool, Int, List, Real, RegLan, String},
    keywordstyle=[30]\color{teal!90!black},
    keywords=[30]{declare-var, %
                  declare-datatype, declare-datatypes, %
                  declare-weight, declare-sort, %
                  define-fun, define-sort},
    keywordstyle=[40]\color{cyan!45!black},
    keywords=[40]{!, \_, Constant, Variable},
    keywordstyle=[50]\color{red!70!darkgray},
    keywords=[50]{assume, %
                  chc-constraint, constraint, inv-constraint},
    keywordstyle=[60]\color{green!45!black},
    keywords=[60]{synth-fun},
    keywordstyle=[70]\color{yellow!30!black},
    keywords=[70]{oracle-assume, oracle-constraint, %
                  declare-oracle-fun, %
                  declare-correctness-oracle, declare-correctness-cex-oracle, %
                  oracle-constraint-cex, oracle-constraint-io, %
                  oracle-constraint-membership, %
                  oracle-constraint-poswitness, oracle-constraint-negwitness},
    keywordstyle=[80]\color{black},
    keywords=[80]{:fwd-decls, :grammars, :oracles, :recursion, :weights},
    keywordstyle=[90]\color{orange!70!black},
    keywords=[90]{fail, infeasible}
}

\lstdefinelanguage{SyGuS-Desugar}[]{SyGuS}{
    basicstyle=\linespread{0.95}\ttfamily,
    numbers=none,
    xleftmargin=0em
}

\global\long\def\code{\lstinline[language=SyGuS, basicstyle=\ttfamily\color{black!80!white}]}

%% file: macros.tex
\theoremstyle{definition}
\newtheorem{example}{Example}
\newcommand{\rem}[1]{\textcolor{red}{[#1]}}
\newcommand{\ajr}[1]{\rem{#1 --ajr}}
\newtheorem{definition}{Definition}

\global\long\def\autobox#1{#1}

\global\long\def\opname#1{\autobox{\operatorname{#1}}}

\global\long\def\dontcare{\_}

\global\long\def\bbracket#1{\left\llbracket #1\right\rrbracket }

\global\long\def\cnot#1{\centernot#1}

\global\long\def\roset#1{\left\{  #1\right\}  }

\global\long\def\ruset#1#2{\roset{#1\;\middle\vert\;#2}}

\global\long\def\union#1#2{#1\cup#2}

\global\long\def\bigunion#1#2{\bigcup_{#1}#2}

\global\long\def\intersection#1#2{#1\cap#2}

\global\long\def\bigintersection#1#2{\bigcap_{#1}#2}

\global\long\def\bigland#1#2{\bigwedge_{#1}#2}

\global\long\def\powerset#1{2^{#1}}

\global\long\def\cart#1#2{#1\times#2}

\global\long\def\tuple#1{\left(#1\right)}

\global\long\def\quoset#1#2{\left.#1\middle/#2\right.}

\global\long\def\equivclass#1{\left[#1\right]}

\global\long\def\refltransclosure#1{#1^{*}}

\newcommandx\funcapphidden[3][usedefault, addprefix=\global, 1=]{#2#1#3}

\global\long\def\funcapplambda#1#2{\funcapphidden[\,]{#1}{#2}}

\global\long\def\funcapptrad#1#2{\funcapphidden{#1}{\tuple{#2}}}

\global\long\def\funccomp#1#2{#1\circ#2}

\global\long\def\arrow#1#2{#1\to#2}

\global\long\def\func#1#2#3{#1:\arrow{#2}{#3}}

\global\long\def\N{\mathbb{N}}

\global\long\def\Z{\mathbb{Z}}

\global\long\def\Q{\mathbb{Q}}

\global\long\def\R{\mathbb{R}}

\global\long\def\D{\mathbb{D}}

\global\long\def\Zmod#1{\quoset{\Z}{#1\Z}}

\global\long\def\vector#1{\mathbf{#1}}

\global\long\def\dotprod#1#2{#1\cdot#2}

\global\long\def\true{\texttt{true}}

\global\long\def\false{\texttt{false}}

\global\long\def\strempty{\epsilon}
\global\long\def\teq{\approx}
\global\long\def\wsym#1#2{\omega_{#1,#2}}

\global\long\def\wgenerates#1#2#3#4{#1 \mapsto_{#2,#3}^{\ast} #4}

\global\long\def\kstar#1{#1^{*}}
\global\long\def\koption#1{#1^{?}}
\global\long\def\kstarn#1#2{#1^{#2}}

\global\long\def\kplus#1{#1^{+}}

\global\long\def\paren#1{\texttt{(}#1\texttt{)}}

%% file: sygus-macros.tex
\global\long\def\sygus{\left\langle SyGuS\right\rangle }

\global\long\def\cmd{\left\langle Cmd\right\rangle }

\global\long\def\smtcmd{\left\langle SmtCmd\right\rangle }

\global\long\def\oraclecmd{\left\langle OracleCmd\right\rangle }

\global\long\def\setlogiccmd{\left\langle SetLogicCmd\right\rangle }

\global\long\def\sortdefcmd{\left\langle SortDefCmd\right\rangle }

\global\long\def\fundefcmd{\left\langle FunDefCmd\right\rangle }

\global\long\def\vardeclcmd{\left\langle VarDeclCmd\right\rangle }

\global\long\def\fundeclcmd{\left\langle FunDeclCmd\right\rangle }

\global\long\def\synthfuncmd{\left\langle SynthFunCmd\right\rangle }

\global\long\def\constraintcmd{\left\langle ConstraintCmd\right\rangle }
\global\long\def\assumecmd{\left\langle AssumeCmd\right\rangle }

\global\long\def\fundefcmdkwd{\left\langle FunDefCmdKwd\right\rangle }

\global\long\def\checksynthcmd{\left\langle CheckSynthCmd\right\rangle }

\global\long\def\setoptscmd{\left\langle SetOptsCmd\right\rangle }

\global\long\def\symbol{\left\langle Symbol\right\rangle }

\global\long\def\identifier{\left\langle Identifier\right\rangle }
\global\long\def\keyword{\left\langle Keyword\right\rangle }
\global\long\def\identifierindex{\left\langle Index\right\rangle }
\global\long\def\qualidentifier{\left\langle QualIdentifier\right\rangle }
\global\long\def\attribute{\left\langle Attribute\right\rangle }
\global\long\def\attributevalue{\left\langle AttributeValue\right\rangle }

\global\long\def\quotedliteral{\left\langle QuotedLiteral\right\rangle }

\global\long\def\sortexpr{\left\langle Sort\right\rangle }

\global\long\def\intconst{\left\langle Numeral\right\rangle }

\global\long\def\realconst{\left\langle Decimal\right\rangle }

\global\long\def\boolconst{\left\langle BoolConst\right\rangle }

\global\long\def\bvconst{\left\langle BVConst\right\rangle }

\global\long\def\hexconst{\left\langle HexConst\right\rangle }
\global\long\def\binaryconst{\left\langle BinConst\right\rangle }

\global\long\def\stringconst{\left\langle StringConst\right\rangle }

\global\long\def\sortedvar{\left\langle SortedVar \right\rangle }
\global\long\def\varbinding{\left\langle VarBinding \right\rangle }

\global\long\def\enumconst{\left\langle EnumConst\right\rangle }

\global\long\def\literal{\left\langle Literal\right\rangle }

\global\long\def\term{\left\langle Term\right\rangle }

\global\long\def\bfterm{\left\langle BfTerm\right\rangle }

\global\long\def\gterm{\left\langle GTerm\right\rangle }

\global\long\def\letterm{\left\langle LetTerm\right\rangle }

\global\long\def\letgterm{\left\langle LetGTerm\right\rangle }

\global\long\def\ntdef{\left\langle GroupedRuleList\right\rangle }

\global\long\def\grammardef{\left\langle GrammarDef\right\rangle }

\global\long\def\feature{\left\langle Feature\right\rangle }

\global\long\def\dtdecl{\left\langle DTDecl\right\rangle }
\global\long\def\dtconsdecl{\left\langle DTConsDecl\right\rangle }

\global\long\def\sortdecl{\left\langle SortDecl\right\rangle }

\global\long\def\option{\left\langle Option\right\rangle }

\global\long\def\fundec{\left\langle FunDec\right\rangle }

\global\long\def\fundef{\left\langle FunDef\right\rangle }


\global\long\def\dtdeclkwd{\texttt{declare-datatype}}

\global\long\def\dtsdeclkwd{\texttt{declare-datatypes}}

\global\long\def\recfundefkwd{\texttt{define-fun-rec}}

\global\long\def\recfunsdefkwd{\texttt{define-funs-rec}}

\global\long\def\setlogickwd{\texttt{set-logic}}
\global\long\def\setinfokwd{\texttt{set-info}}

\global\long\def\sortdefkwd{\texttt{define-sort}}

\global\long\def\sortdeclkwd{\texttt{declare-sort}}

\global\long\def\fundefkwd{\texttt{define-fun}}

\global\long\def\vardeclkwd{\texttt{declare-var}}

\global\long\def\primedvardeclkwd{\texttt{declare-primed-var}}

\global\long\def\fundeclkwd{\texttt{declare-fun}}

\global\long\def\synthfunkwd{\texttt{synth-fun}}

\global\long\def\synthinvkwd{\texttt{synth-inv}}

\global\long\def\oracledeckwd{\texttt{declare-oracle-fun}}

\global\long\def\iooracledeckwd{\texttt{oracle-constraint-io}}
\global\long\def\cexoracledeckwd{\texttt{oracle-constraint-cex}}
\global\long\def\corroracledeckwd{\texttt{declare-correctness-oracle}}
\global\long\def\memoracledeckwd{\texttt{oracle-constraint-membership}}
\global\long\def\corrcexoracledeckwd{\texttt{declare-correctness-cex-oracle}}
\global\long\def\poswitnessoracledeckwd{\texttt{oracle-constraint-poswitness}}
\global\long\def\negwitnessoracledeckwd{\texttt{oracle-constraint-negwitness}}

\global\long\def\weightdeclkwd{\texttt{declare-weight}}


\global\long\def\constraintkwd{\texttt{constraint}}
\global\long\def\assumekwd{\texttt{assume}}
\global\long\def\constraintinvkwd{\texttt{inv-constraint}}
\global\long\def\constraintchckwd{\texttt{chc-constraint}}

\global\long\def\constraintoraclekwd{\texttt{oracle-constraint}}
\global\long\def\assumeoraclekwd{\texttt{oracle-assume}}

\global\long\def\checksynthkwd{\texttt{check-synth}}
\global\long\def\optsynthkwd{\texttt{optimize-synth}}

\global\long\def\setoptkwd{\texttt{set-option}}

\global\long\def\setoptskwd{\texttt{set-options}}

\global\long\def\bitveckwd{\texttt{BitVec}}

\global\long\def\arraykwd{\texttt{Array}}

\global\long\def\intkwd{\texttt{Int}}

\global\long\def\boolkwd{\texttt{Bool}}

\global\long\def\enumkwd{\texttt{Enum}}

\global\long\def\realkwd{\texttt{Real}}

\global\long\def\constantkwd{\texttt{Constant}}

\global\long\def\varkwd{\texttt{Variable}}

\global\long\def\inputvarkwd{\texttt{InputVariable}}

\global\long\def\localvarkwd{\texttt{LocalVariable}}

\global\long\def\letkwd{\texttt{let}}

\global\long\def\funsortkwd{\texttt{->}}
\global\long\def\implopkwd{\texttt{=>}}

\global\long\def\askwd{\texttt{as}}

\global\long\def\forallkwd{\texttt{forall}}
\global\long\def\existskwd{\texttt{exists}}

\global\long\def\truekwd{\texttt{true}}
\global\long\def\falsekwd{\texttt{false}}

\global\long\def\setfeaturekwd{\texttt{set-feature}}

\global\long\def\setfeaturesetkwd{\texttt{set-feature-set}}

\global\long\def\fgrammarskwd{\texttt{grammars}}
\global\long\def\fquantifierskwd{\texttt{quantifiers}}
\global\long\def\ffwddeclskwd{\texttt{fwd-decls}}
\global\long\def\frecursionkwd{\texttt{recursion}}
\global\long\def\foracleskwd{\texttt{oracles}}
\global\long\def\fweightskwd{\texttt{weights}}

\global\long\def\sbool{\mathsf{Bool}}

\global\long\def\slogic{\mathcal{L}}

\global\long\def\kweight{\texttt{:weight}}
\global\long\def\kweights#1{\texttt{:weight.#1}}
\global\long\def\weightsymnf#1#2{\texttt{(\_\ }#1\ #2\texttt{)}}
\global\long\def\weightsym#1#2{\texttt{(\_\ #1\ #2)}}